\documentstyle[aps,preprint,epsfig]{revtex}
\newcommand{\bq}{\begin{equation}}
\newcommand{\eq}{\end{equation}}
\newcommand{\bqn}{\begin{eqnarray}}
\newcommand{\eqn}{\end{eqnarray}}
\newcommand{\nb}{\nonumber}
\newcommand{\lb}{\label}
\begin{document}
 
\title{Gravitational collapse of spherically symmetric perfect fluid with
kinematic self-similarity}
\author{C.F.C. Brandt \thanks{E-mail  address:
fred@dft.if.uerj.br} ${ }^{a}$, 
L.-M. Lin \thanks{E-mail  address: lmlin@artsci.wustl.edu} ${ }^{b}$,
J.F. Villas da Rocha \thanks{E-mail  address:
roch@dft.if.uerj.br} ${ }^{c}$, 
\ and A.Z. Wang \thanks{E-mail  address:
wang@dft.if.uerj.br} ${ }^{c}$}
\address{${ }^{a}$ Departamento de Astronomia  Gal\'atica e Extra-Gal\'atica, 
Observat\'orio Nacional, Rua General Jos\'e Cristino 77, S\~ao
Cristov\~ao, 20921-400 Rio de Janeiro~--~RJ, Brazil\\
${ }^{b}$ McDonnell Center for the Space Sciences, Department of Physics,
Washington
University, St. Louis,  Missouri 63130, USA\\
${ }^{c}$ Departamento de F\' {\i}sica Te\' orica, 
Universidade do Estado do Rio de Janeiro, Rua S\~ ao
 Francisco Xavier $524$, Maracan\~ a, $20550-013 $
  Rio de Janeiro~--~RJ, Brazil}
  
\date{May 3, 2001}

\maketitle

\newpage

\begin{abstract}

Analytic spherically symmetric solutions of the Einstein field equations
coupled with a perfect fluid and with self-similarities of the zeroth,
first and second kinds, found recently by Benoit and Coley [Class. Quantum
Grav. {\bf 15}, 2397 (1998)], are studied, and found that some of them
represent   gravitational collapse. When the solutions
have   self-similarity of the first (homothetic) kind, some of the solutions
may represent critical collapse but in the sense that now
the ``critical" solution separates the collapse that forms black
holes from the collapse that forms naked singularities.  The formation of
such  black holes {\em always} starts with a mass gap,
although   the ``critical" solution has homothetic self-similarity. The
solutions with   self-similarity of the zeroth and second kinds seem
irrelevant to  critical collapse.  Yet, it is also found that 
the de Sitter solution is a particular case of the solutions with
self-similarity of the zeroth kind, and that the Schwarzschild solution is a
particular case of  the solutions with self-similarity of the second kind
with the index $\alpha = 3/2$.

\end{abstract}
 
%%%%%%%%%%%%%%%%%%%%%%%%%%%%%%%%%%%%%%%%%%%%%%%%%%%%%%%%%%%

\section{Introduction}

\renewcommand{\theequation}{1.\arabic{equation}}
\setcounter{equation}{0}
 
Gravitational collapse of a realistic body has been one of the most important
and thorny subjects in General Relativity (GR) since the  early times of
GR   \cite{Joshi93}. Quite recently, thanks to Choptuik's numerical discovery
of critical phenomena in the threshold of black hole formation \cite{Ch93}, the
subject has attracted further attention. As a matter of fact, it is so
attractive that {\em Critical Phenomena in Gravitational Collapse} has been
already a very established sub-area in GR \cite{Gu00,Wang01}. 
From all the work done so far,
the following seems clear: (a) The critical solution and the two
dimensionless constants $\triangle$ and $\gamma$   are universal only with
respect to  the same matter field, and  usually are matter-dependent.
(b) The universality of the critical
solution and the exponent $\gamma$ now  are well understood in terms of
perturbations \cite{HKA96},   while the physical origin  of   $\triangle$ still
remains somewhat of a mystery. The former is   closely related to the fact
that the critical solution has only one unstable  mode. This property now is
considered as the main criterion for a solution to  be critical.
(c) The critical solutions can have discrete self-similarity (DSS) or
homothetic self-similarity  (HSS) \footnote{In the literature,  
homothetic self-similarity has been also called continuous self-similarity. 
However, in order to distinguish it from the self-similarity of
the other kinds, in this paper 
we shall refer it as homothetic self-similarity, or self-similarity of the 
first kind.}, or none of them, depending on the matter  
fields and regions of  the initial data spaces.   So far, in all
the cases  where the critical solution either has DSS or HSS, the formation of
black holes {\em always} turns on with zero mass, the so-called Type II
collapse, while in the cases in which the critical solution has neither DSS
nor HSS, the formation {\em always} turns on with a mass gap,  the so-called
Type I collapse \cite{Gu00,Wang01}. 

In the Type II collapse, it is the usual belief that the fact
that black hole starts to form with an infinitesimal mass is closely related
to the fact that the problem concerned is of scale invariance, for example, 
the Einstein equations coupled with a massless scalar
field. When the scalar field is massive, the
corresponding field equations are scale-invariant only asymptotically
\cite{Gu00}. For a perfect fluid with the equation of state $ p = k \rho$, the
corresponding Einstein field equations are also of scale invariance. 
As a result, in all these cases critical phenomena of Type II collapse were 
found, and in the case of the scalar field the critical solution has DSS,
while in the case of perfect fluid, the critical solutions have
HSS \cite{Ch93,BCG97,EC94}. 

It is known that homothetic self-similarity is a particular case of
kinematic  self-similarity \cite{CH89}. In fact, the latter consists of three
kinds, the zeroth kind, the first (homothetic) kind, and the second kind.
Thus, a natural  question is: Can critical solutions have
self-similarity of the  other kinds? 

In this paper, we shall study this problem for the gravitational  
collapse of perfect fluid with kinematic self-similarity. As a matter of fact, 
several classes of such analytic solutions to the Einstein field equations 
are already known \cite{BC98}. So, here we shall study these solutions
in some details and pay particular attention on critical solutions. Finding
critical solutions usually consists of two steps, one is first to find a
generic family (or families) of solutions,  
characterized, say, by a parameter $p$, such that when 
$p > p^{*}$ the collapse forms black holes, and when $p < p^{*}$ it does not.
Once such solutions are found, one needs to make perturbations of the
solution $p = p^{*}$ and to study the spectrum of their modes. If
the solution  has only one unstable mode, then by definition this solution is
 a critical solution, and the exponent $\gamma$ is given by
\bq
\lb{exponent}
\gamma = \frac{1}{|\sigma_{1}|},
\eq
where $\sigma_{1}$ is the unstable mode \cite{HKA96}. In this paper, we shall
consider only  the first part of the problem, and leave the study of
perturbations to another occasion.  Specifically,  the paper is organized as
follows: In Sec. II  we shall give a brief introduction to kinematic
self-similarity, and in Sec. III we shall study the  Benoit-Coley (BC)
solutions with self-similarity of the zeroth kind, while in Sec. IV   the
BC solutions with self-similarity of the first and second   kinds will be
studied.   The paper is closed with Sec. V, in which our main conclusions  are
presented.  An appendix is also included, where the Einstein field equations
are written in terms of self-similar variables.  

\section{Spherically Symmetric Spacetimes with Kinematic Self-Similarity}

\renewcommand{\theequation}{2.\arabic{equation}}
\setcounter{equation}{0}

Self-similarity
refers to the fact that the spatial distribution of the characteristics of
motion remains similar to itself at all times in which all dimensional
constant parameters entering the initial and boundary conditions vanish or
become infinite \cite{BZ72}. Such solutions describe the ``intermediate
asymptotic" behavior of solutions in the region where a solution no longer
depends on the details of the initial and$/$or boundary conditions. 

Cases in which the form of the self-similar
asymptotes can be obtained from dimensional considerations are referred to as
self-similarity of the {\em first (homothetic) kind} \cite{BZ72}. Solutions of
the first kind were first studied by Cahill and Taub  in GR for a perfect fluid
\cite{CT71}.  They showed that the existence of  self-similarity (of the
first kind) could be formulated invariantly in terms of a homothetic Killing
vector, $\xi^{\mu}$, which satisfies the conformal Killing equation,
\bq
\lb{1.1}
{\cal{L}}_{\xi}g_{\mu\nu} = 2 g_{\mu\nu},
\eq
where ${\cal{L}}$ denotes Lie differentiation along $\xi^{\mu}$. From the
above it can be shown that 
\bq
\lb{1.2}
{\cal{L}}_{\xi}G_{\mu\nu} = 0.
\eq
 For a perfect fluid with the energy-momentum tensor (EMT) given by
\bq
\lb{1.3}
T_{\mu\nu} = (p + \rho)u_{\mu}u_{\nu} - p g_{\mu\nu},
\eq
it can be shown that it is consistent with Eq.(\ref{1.2}) if we require
\bq
\lb{1.4}
{\cal{L}}_{\xi}u^{\mu} = - u^{\mu},\;\;\;\;
{\cal{L}}_{\xi}\rho = - 2\rho,\;\;\;\;
{\cal{L}}_{\xi} p = - 2p.
\eq
Hence, in this case ``geometric" self-similarity and ``physical" similarity
coincide, although this does not need to be so in more general cases
\cite{CH89,Coley97}. 
Applying the above to the spacetimes with spherical symmetry,
\bq 
\lb{1.5}
ds^{2} = r^{2}_{1}\left\{e^{2\Phi(t, r)}dt^{2} - e^{2\Psi(t, r)}dr^{2} 
- r^{2}S^{2}(t, r)d\Omega^{2}\right\}, 
\eq
where $d\Omega^{2} = d\theta^{2} + \sin^{2}\theta d\varphi^{2}$, 
Cahill and Taub found that the condition (\ref{1.1}) requires
\bq
\lb{1.6}
\Phi(t, r) = \Phi(\xi),\;\;\;
\Psi(t, r) = \Psi(\xi),\;\;\;
S(t,r) = S(\xi), 
\eq
where  
\bq
\lb{1.7}
\xi = \frac{r}{-t}.
\eq
The corresponding homothetic Killing vector $\xi^{\mu}$ is given by
\bq
\lb{1.8}
\xi^{\mu}\frac{\partial}{\partial x^{\mu}} = t\frac{\partial}{\partial t} +  r
\frac{\partial}{\partial r}. 
\eq

Note that in writing the metric (\ref{1.5}) we had multiplied  a factor
$r^{2}_{1}$ to the usual spherical metric, so that the metric coefficients
$\Phi, \; \Psi,\; S$ and the coordinates $t, \; r, \; \theta$ and $\varphi$
now are all {\em dimensionless}, where we assume that $r_{1}$ has the dimension
of length. It is found that this choice will simplify  the dimensional
analysis to be given  below. The corresponding Einstein tensor and Einstein
field equations are given in terms of both $t, \; r$ and $x, \; \tau$ in the
Appendix, where $x$ and $\tau$ are the self-similar variables that are
functions of $t$ and $r$. Their explicit definitions in each case are given in
the Appendix.

 The existence of self-similarity of the first kind is closely
related to the conservation laws and to the invariance of the problem with
respect to the group of similarity transformations of quantities with
independent dimensions, in which case a certain regularity of the limiting
process in passing from original non-self-similar regime to the self-similar
regime is assumed implicitly. However, in general such a passage does not need
to be regular.  Consequently, the expressions for the self-similar variables
are not determined from  dimensional analysis. Such solutions are then called
self-similar solutions of the {\em second kind}. A characteristic of these
solutions is that they {\em contain dimensional constants} that are not
determined from the conservation laws \cite{BZ72}.
Using these arguments to a perfect fluid (\ref{1.3}), Carter and Henriksen
\cite{CH89} gave the notion of {\em kinematic self-similarity} with its
properties,
\bq
\lb{1.9}
{\cal{L}}_{\xi}h_{\mu\nu} = 2 h_{\mu\nu},\;\;\;\;
{\cal{L}}_{\xi}u^{\mu} = -\alpha  u^{\mu},
\eq
where $h_{\mu\nu}$ is the project operator, defined by
\bq
\lb{1.10}
h_{\mu\nu} = g_{\mu\nu} - u_{\mu} u_{\nu},
\eq
and $\alpha$ is an arbitrary {\em dimensionless} constant. When $\alpha = 1$,
it can be shown that the kinematic self-similarity reduces to the
self-similarity of the first kind (homothetic self-similarity). When $\alpha
\not= 1$, Carter and Henriksen argued that this would be a natural
relativistic counterpart of self-similarity of the second kind ($\alpha \not=
1$), and of  the zeroth kind ($\alpha = 0$), in Newtonian Mechanics. 

Applying the above to the
spherical case, Carter and Henriksen found that the metric coefficients
$\Phi,\; \Psi$ and $S$ should also take the form of Eq.(\ref{1.6}) but with
the self-similar variable $\xi$ and conformal vector $\xi^{\mu}$ now being
given, respectively,  by 
\bq 
\lb{1.11}
\xi^{\mu}\frac{\partial}{\partial x^{\mu}} = \alpha t\frac{\partial}{\partial
t} +  r \frac{\partial}{\partial r},\;\;\;
\xi = \frac{r}{\left(- t\right)^{1/\alpha}},\; (\alpha \not= 0),
\eq
for the second kind, and 
\bq
\lb{1.12}
\xi^{\mu}\frac{\partial}{\partial x^{\mu}} = \frac{\partial}{\partial t} +
 r \frac{\partial}{\partial r},\;\;\;
\xi = {r}e^{-t},\; (\alpha = 0),
\eq
for the zeroth kind. Comparing Eq.(\ref{1.8}) with Eq.(\ref{1.11}) we find
that the self-similarity of the first kind can be considered as a particular
case of the one of the second kind. In this paper we shall do so, although, as
we mentioned above, the physics is quite different in the two
cases. In particular, when the coordinates $t$ and $r$ are rescaled, $t' = ct$
and $r' = cr$, where $c$ is a constant, $\xi$ is unchanged only for the
homothetic case $\alpha = 1$.

\section{Analytic Solutions of Perfect Fluid with  Self-Similarity of the Zeroth
Kind and Their Physical Interpretations}

\renewcommand{\theequation}{3.\arabic{equation}}
\setcounter{equation}{0}

The solutions  to be studied in this section are those  given by Eqs.(2.54) -
(2.57) in \cite{BC98}. Note that the expression for the function $S(x)$
given there is not correct. As a matter of fact, setting  $\Phi = 0$, we find
that Eq.(\ref{B.11b}) yields 
\bq 
\lb{2.1} 
e^{2\Psi} = (1+y)^{2}S^{2},
\eq
while Eq.(\ref{B.9}) is satisfied automatically. Submitted Eq.(\ref{2.1}) into
Eq.(\ref{B.11a}), we obtain 
\bq
\lb{2.1a}
y_{,xx} + 3yy_{,x} =0,
\eq
which has the first integral
\bq
\lb{2.4}
2y_{,x} + 3y^{2} + p_{0} = 0,
\eq
where $x = \ln(\xi)$, and $p_{0}$ is a {\em dimensionless} constant, in
contrast to the claim given  in \cite{BC98}.  Then, it can be shown that the
corresponding perfect fluid is given by \footnote{In this paper the units will
be chosen such that the Einstein coupling constant $ \kappa \left[\equiv 8\pi
G/c^{4} \right] = 1.$} 
\bq
\lb{2.3}
\rho = \frac{y(3y - p_{0})}{r^{2}_{1}(1 + y)}, \;\;\; 
p = \frac{p_{0}}{r^{2}_{1}},\;\;\; u_{\mu} = r_{1}\delta^{t}_{\mu},
\eq
where  $ y(x) \equiv S_{,x}/{S}$.
Depending on the sign of $p_{0}$, Eq.(\ref{2.4}) has physically 
different solutions.  In the following let us consider them separately.

\subsection{ Case $p_{0} = 0$}

In this case, it can be shown that Eq.(\ref{2.4}) has the
solution, 
\bq
\lb{2.6}
y(x) = \frac{2}{3(x + x_{0})}, \;\;\;\; 
S(x) = S_{0}(x + x_{0})^{2/3},
\eq
where $S_{0}$ and $x_{0}$  are integration constants. Without loss of
generality, we can set $S_{0}$ equal to one by a conformal
transformation. In the following we shall assume that this is
always done whenever it is applicable. Substituting the above 
expressions into Eq. (\ref{2.3}), we
find that  
\bq
\lb{2.7}
\rho = \frac{4}{r^{2}_{1}(x + x_{0})\left[3(x + x_{0}) + 2\right]},\;\;\;\;
p = 0,
\eq
which shows that in this case the solutions represent a dust fluid. Thus, these
solutions  must belong to the general Tolmann-Bondi class 
 \cite{TB34}.  

\begin{figure}[htbp]
\begin{center}
\leavevmode
    \epsfig{file=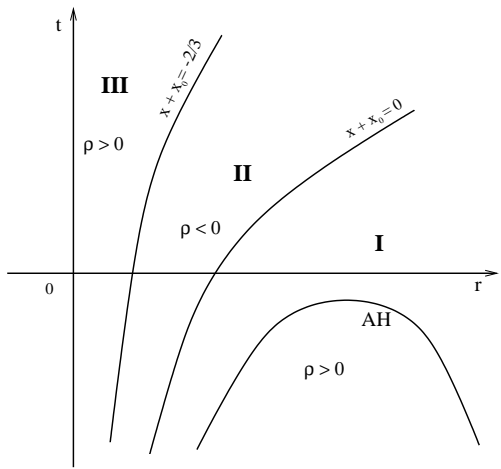,width=0.4\textwidth,angle=0}
\caption{The spacetime described by the solutions of Eq.(\ref{2.6}) in the
text in the ($t, r$)-plane. It is singular on the hypersurfaces $x + x_{0} =
0$  and $x + x_{0} = - 2/3$, which divide the whole spacetime into three
disconnected regions:   $I = \left\{x^{\mu}: x + x_{0} \ge
0\right\},\;  II = \left\{x^{\mu}: -2/3 \le x + x_{0} \le 0\right\}$, and $
III = \left\{x^{\mu}: x + x_{0} \le -2/3\right\}$.}
    \label{fig1}
  \end{center}
\end{figure}

\noindent From Eq.(\ref{2.7}) we can see that the
spacetime is singular on the hypersurfaces $x + x_{0} = 0$ and $x + x_{0} = -
2/3$. These two hypersurfaces divide the whole spacetime into three regions,
I, II, and III, where $I = \left\{x^{\mu}: x + x_{0} \ge 0\right\},\;  II =
\left\{x^{\mu}: -2/3 \le x + x_{0} \le 0\right\}$, and $ III = \left\{x^{\mu}:
x + x_{0} \le -2/3\right\}$ [See  Fig. 1]. In region II, the energy
density of the fluid $\rho$ is negative, and the physics of the spacetime in
this region is not clear. In region III,    it is non-negative and the
singularity located on the hypersurface $x + x_{0} = - 2/3$ is naked, and the
spacetime in this region can be considered as representing an inhomogeneous
cosmological model. In region I, to
study the nature of the singularity located on the hypersurface $x + x_{0} =
0$, let us first calculate the gradient of the geometric radius, $R \equiv r
S(x)$, of the two sphere,   
 \bq
\lb{2.8}
R_{,\alpha}R_{,\beta} g^{\alpha\beta} = \frac{1}{9(x +
x_{0})^{2/3}}\left\{4 r^{2} - 9(x + x_{0})^{2/3}\right\}.
\eq
The formation of apparent horizons are indicated by the vanishing of
the gradient. Thus, setting the right-hand side of Eq.(\ref{2.8}) to zero, we
obtain 
\bq
\lb{2.9}
x + x_{0} = \left(\frac{2r}{3}\right)^{3}, 
\; \left(R_{,\alpha}R_{,\beta} g^{\alpha\beta} = 0\right).
\eq
Clearly, for any given  $r$, we always have $x + x_{0} \ge 0$ on the
apparent horizon. Hence, in the present case the formation of the spacetime
singularity on the hypersurface $x + x_{0} = 0$ always follows the formation
of the apparent horizon, or in other words, the singularity is always covered
by the apparent horizon. These
solutions can be considered as representing the formation of black holes due
to the gravitational collapse of the fluid, starting at the moment $t = -
\infty$. Defining the mass function $m(t,r)$ as \cite{PI90}, 
\bq
\lb{2.10}
m(t, r) = \frac{R_{}}{2}\left(1 + R_{,\alpha}R_{,\beta}
g^{\alpha\beta}\right), 
\eq    
we find that $m(t, r) = {2r^{3}}/{9}$. On the apparent horizon, we have
\bq
\lb{2.11a}
M_{BH}(t) = m\left(t, r_{AH}(t)\right) = \frac{2r^{3}_{AH}(t)}{9},
\eq
where $r_{AH}(t)$ is a solution of Eq.(\ref{2.9}). The quantity $M_{BH}(t)$
can be considered as the contribution of the collapsing perfect fluid to
the total mass of such   formed black holes \cite{Gu00}, which in the
present case goes to infinity as $r_{AH}(t) \rightarrow + \infty$, as can be
seen from Eq.(\ref{2.9}). Therefore, now the collapse of the
dust fluid always forms black holes with infinitely large mass.  To remedy this
shortage, one may cut the spacetime along a non-spacelike hypersurface, say,
$r = r_{0}(t)$, and then join the region, $r \le r_{0}(t)$, to an
asymptotically flat region  \cite{WO97}. By this way, we can see that the
resultant model will represent the collapse of a dust ball with its radius $ r
= r_{0}(t)$ [cf. Fig. 2]. From the moment $t = t_{f}$ on, the ball collapses
completely inside the apparent horizon, and the contribution of the collapsing
ball to the total mass of such a formed black hole is given by  \bq \lb{2.12}
M_{BH} = M_{BH}\left(t_{f}\right),  
\eq
where $t_{f}$ is a solution of the equation,
\bq
\lb{2.13}
r_{0}(t_{f}) = r_{AH}(t_{f}).
\eq
In the present case, since the fluid is co-moving with the coordinates, without
loss of generality, we can choose the joining hypersurface as $r_{0}(t) =
r_{0} = Const.$ Then, from Eq.(\ref{2.12}) we find that $M_{BH} (=
2r^{3}_{1}/9)$ is always finite and different from zero for any given non-zero
$r_{0}$. This is different from the gravitational collapse of dust fluid with
 self-similarity of the first kind  studied in \cite{WRS97}, where it was shown
 that for any given non-zero $r_{0}$, black holes with infinitesimal
mass can be formed by properly choosing a parameter that characterizes the
strength of the collapse.  

\vspace{1.cm}

\begin{figure}[htbp]
  \begin{center}
    \leavevmode
    \epsfig{file=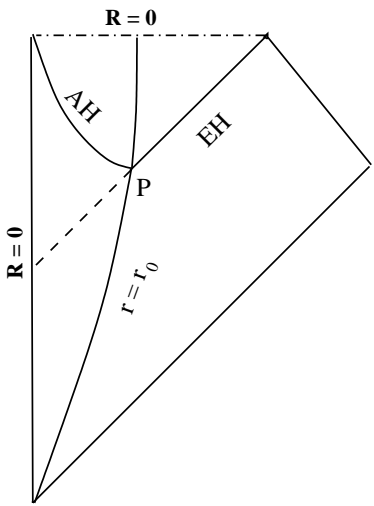,width=0.3\textwidth,angle=0}
    \caption{ The corresponding Penrose diagram of the spacetime described by
 Eq.(\ref{2.6}), after it is first cut along the hypersurface $r = r_{0}$ and
then joined with an asymptotically flat region. The point $P$ represents the
moment when $r_{0} = r_{AH}(t_{f})$, where the ball of the fluid collapses
completely inside the apparent horizon.}
    \label{fig2}
  \end{center}
\end{figure}

\subsection{Case $p_{0} > 0$}

In this case it can be shown that Eq.(\ref{2.4}) has the
solution,
\bq
\lb{2.14}
y(x) = - a \tan A(x), \;\;\;\;
S(x) =  \cos^{2/3}A(x),
\eq
where   $x_{0}$ is another integration constant, and
\bq
\lb{2.14a}
A(x) \equiv \frac{3a}{2}\left(x + x_{0}\right),\;\;\;\;
a \equiv \left|\frac{p_{0}}{3}\right|^{1/2},
\eq
Then, from Eq.(\ref{2.3}) we find that 
\bq
\lb{2.15}
\rho = \frac{3a^{2} \tan A(x)
\left[\tan A(x) - a\right]}{r^{2}_{1}\left[1 - a
\tan A(x)\right]},\;\;\;\;\;
p = \frac{p_{0}}{r^{2}_{1}} > 0,
\eq
from which we can see that the solutions are singular on the
hypersurfaces,
\bq
\lb{2.16}
x +  x_{0} = \cases{\frac{2}{3a}\left[{\tan^{-1}}\left(\frac{1}{a}\right)  +
n \pi \right],\cr
\frac{2\pi}{3a}\left(n + \frac{1}{2}\right) \cr} \;\; (a \not= 1),
\eq
except for the case $a = 1$, where we have 
\bq
\lb{2.16a}
\rho = - \frac{3}{r^{2}_{1}} \tan\left\{\frac{3}{2}(x + x_{0})\right\}, \;\;\;
(a = 1), 
\eq
where $n$ is an integer. Clearly, in the latter case the spacetime is singular
only on the hypersurfaces
\bq
\lb{2.16b}
x + x_{0} =  \frac{2\pi}{3}\left(n + \frac{1}{2}\right), \;\; (a = 1).
\eq
However, in either of the two cases, the spacetime is singular on various
hypersurfaces, and the energy conditions, weak, strong and dominant
\cite{HE73}, hold only in certain regions. The physics of these singularities
are not clear, and the solutions may have physical applications only in
certain regions. In particular, they cannot be interpreted as representing 
gravitational collapse of the fluid.

\subsection{Case $p_{0} < 0$}

In this case, the solutions can be further classified into three different
cases, according to $\alpha) \; y^{2} > a^{2}, \; \beta) \; y^{2} = a^{2}$,
and  $\gamma) \; y^{2} < a^{2}$.

{\bf{ $\alpha)\;$ Case $\; y^{2} > a^{2}$}}: In this case it can be show
that Eq.(\ref{2.4}) has the solution, 
\bq
\lb{2.17}
y(x) = a \frac{\cosh A(x)}{\sinh A(x)}, \;\;\;\;
S(x) =   \sinh^{2/3}A(x), 
\eq
while Eqs.(\ref{2.3}) yields,
\bqn
\lb{2.18}
\rho &=& \frac{3a^{2}\cosh A(x) \left[\cosh A(x) + a \sinh A(x)\right]}
{r^{2}_{1}\sinh A(x) \left[\sinh A(x) + a \cosh A(x)\right]},\nb\\
p &=& - \frac{|p_{0}|}{r^{2}_{1}} = - \frac{3 a^{2}}{r^{2}_{1}},
\eqn
where $A(x)$ and $a$ are still given by Eq.(\ref{2.14a}).  
The three energy conditions now require $\rho \ge 9 a^{2}$. Then, from
Eq.(\ref{2.18}) we can see that this condition holds only in the
region 
\bq
\lb{2.19}
\ln r + x_{0} - x_{1} \le t \le \ln r + x_{0},
\eq
for any given $a$, or in the region 
\bq
\lb{2.19a}
t \le \ln r + x_{0} + x_{2},\;\; (a < 1), 
\eq
for $a < 1$, where
\bqn
\lb{2.20}
x_{1} &\equiv& \frac{1}{3a}\ln\left(\frac{2 + \sqrt{3 + a^{2}}}{1 + a}\right)
 > 0,\nb\\ 
x_{2} &\equiv& - \frac{1}{3a}\ln\left|\frac{2 - \sqrt{3 +
a^{2}}}{1 + a}\right| > 0. 
\eqn
In the region defined by Eq.(\ref{2.19}),  
the spacetime is limited by the  curvature singularity located at $x + x_{0} =
0$ in one side, and by the hypersurface $x+ x_{0} = x_{1}$ in the other side,
across the latter the energy conditions do not hold. The solutions in this
region
seem not to have much physics. In the region defined by Eq.(\ref{2.19a}),
the spacetime may be considered as
representing a cosmological model. It is interesting to
note that the spacetime in this region is free of singularities and
asymptotically flat as $ t \rightarrow + \infty$. However, it is not
geodesically
complete and needs to be extended beyond the hypersurface $x + x_{0} = - x_{2}$.
A ``natural" extension would be the one simply given by the above
solutions (\ref{2.17}). This extension will be valid until the hypersurface
$x + x_{0} = - x_{3}$, where 
\bq
\lb{2.21}
x_{3} \equiv - \frac{1}{3a}\ln\left|\frac{1 - a}{1 + a}\right|,
\eq
on which the spacetime is singular. Obviously, the fluid in this extended
region do not satisfy all the three energy conditions.   

{\bf $\beta)\;$ Case $\; y^{2} = a^{2}$}: In this case, it can be shown that
Eq.(\ref{2.4}) has the solution, 
\bq
\lb{2.22}
y(x) = \pm a, \;\;\;\;
S(x) =   e^{\pm a x}, \;\;\;\;
\rho = - p = \frac{3 a^{2}}{r^{2}_{1}}.  
\eq  
Introducing a new radial coordinate $\bar{r}$ via the relation, $\bar{r} =
r^{1\pm a}$, we find that the corresponding metric can be written in the form,
\bq
\lb{2.23}
ds^{2} = r^{2}_{1}\left\{dt^{2} - e^{\pm 2 a (x_{0} - t)}\left(d\bar{r}^{2} +
\bar{r}^{2}d^{2}\Omega\right)\right\},
\eq
which is the de Sitter solution \cite{HE73}.

{\bf $\gamma)\;$ Case $\; y^{2} < a^{2}$}: In this case, we find that 
\bqn
\lb{2.24}
y(x) &=& a \frac{\sinh A(x)}{\cosh A(x)}, \;\;\;\;
S(x) =   \cosh^{2/3}A(x), \nb\\
\rho &=& \frac{3a^{2}\sinh A(x) \left[\sinh A(x) + a \cosh A(x)\right]}
{r^{2}_{1}\cosh A(x) \left[\cosh A(x) + a \sinh A(x)\right]},\nb\\
p &=& - \frac{3 a^{2}}{r^{2}_{1}}.
\eqn
It can be shown that now the energy conditions hold only in the
region,
\bq
\lb{2.25}
\ln r + x_{0} + x_{3} \le t \le \ln r + x_{0} + x_{4}, \; (a \ge 1),
\eq
for $a \ge 1$, where $x_{3}$ is  given by Eq.(\ref{2.21}) and $x_{4}$ is
given by
\bq
\lb{2.25a}
x_{4} \equiv - \frac{1}{3a}\ln \left|\frac{a -1 }{2 + \sqrt{3 + a^{2}}}\right|.
\eq
When $a < 1$, there does not exist any region in which the three energy
conditions hold. The spacetime is singular on the hypersurface $x + x_{0} = -
x_{3}$. The physics of the spacetime in this case is not clear (if there is
any).

\section{Analytic Solutions of Perfect Fluid with   Self-Similarity of the
First and Second Kinds and Their Physical Interpretations}

\renewcommand{\theequation}{4.\arabic{equation}}
\setcounter{equation}{0}

The solutions  to be studied in this section are those  given by Eqs.(2.27) -
(2.31) in \cite{BC98}, for which we have $\Phi = 0$. From Eq.(\ref{C.8b}) we
find that the function $\Psi$ takes the same form as that given by
Eq.(\ref{2.1}) in terms of $y$, while Eq.(\ref{C.8a}) yields,  
\bq
\lb{eq3.2a}
y_{,xx} + (3y + \alpha) y_{,x} =0,
\eq
which allows the first integral,
\bq
\lb{3.2}
2y_{,x} + 3y^{2} + 2\alpha y + \alpha^{2}p_{0} = 0,
\eq
where $p_{0}$ is also a {\em dimensionless} constant, in contrast to what
claimed in \cite{BC98}.  It can be shown that in the present case
Eq.(\ref{C.5}) is satisfied automatically, too. Then,   the corresponding 
perfect fluid  is given by 
\bq 
\lb{3.3}
\rho = \frac{y\left[(3 - 2\alpha)y -
\alpha^{2}p_{0}\right]}{\alpha^{2}r^{2}_{1}(1 + y)t^{2}}, \;\;\;\;
p = \frac{p_{0}}{r^{2}_{1}t^{2}},\;\;\; u_{\mu} = r_{1}\delta^{t}_{\mu}.
\eq
When $\alpha = 1$, the corresponding solutions have   
self-similarity of the first kind, otherwise, they have the
second kind. 

Note that the solutions of Eq.(\ref{3.2}) for the function $y(x)$ given in
\cite{BC98} are  not correct. Thus, in the following we shall first derive the
correct expressions for $y(x)$ and $S(x)$, and then study the physics of the
solutions.  Depending on the value of $p_{0}$, the solutions of Eq.(\ref{3.2})
can be divided into several classes. In the following let us consider them one
by one.

\subsection{Case $p_{0} = 0$}

When $p_{0} = 0$, Eq.(\ref{3.2}) has the solution,
\bqn
\lb{3a.1}
y(x) &=& \frac{2\alpha}{3\left[e^{\alpha(x+x_{0})} - 1\right]},\nb\\
S(x) &=& e^{-\frac{2\alpha}{3}(x+x_{0})}
\left[e^{\alpha(x+x_{0})} - 1\right]^{2/3},
\eqn
where $x_{0}$ is another integration constant. The corresponding energy
density of the fluid is given by
\bq
\lb{3a.2}
\rho = \frac{4(3 -
2\alpha)}{9r^{2}_{1}}\left\{t^{2}\left[e^{\alpha(x+x_{0})} - 1\right]
 \left[e^{\alpha(x+x_{0})} + \frac{2\alpha -
3}{3}\right]\right\}^{-1}. 
\eq
Since $p = 0$ in the present case, the above solutions must also belong to the
general Tolmann-Bondi solutions \cite{TB34}.

{\bf a) Case $\;  0 < \alpha < 1$}: In this subcase, Eq.(\ref{3a.2})
shows that the spacetime is singular on the hypersurfaces,
\bq
\lb{3a.3}
a)\;\; t = 0,\;\;\;\;\;
b)\;\; x + x_{0} = 0,\;\;\;\;\;
c)\;\; x + x_{0} = - x_{5}, 
\eq
where $x_{5}$ is defined as
\bq
\lb{3a.4}
x_{5} \equiv - \frac{1}{\alpha}\ln\left|\frac{3- 2\alpha}{3}\right| > 0.
\eq
From the above we can show that $\rho$ is non-negative only in the region $x
+ x_{0} \le - x_{5}$ or in the region $x + x_{0} \ge 0$. In the region $x +
x_{0} \ge 0$, the spacetime is singular on the two hypersurfaces $x + x_{0} =
0$ and $t = 0$. While the physical meaning of the spacetime in the region $x +
x_{0} \ge 0$, and $ t \le 0$ is not clear, the spacetime in the region $t \ge
0$ can be considered as representing a cosmological model with its initial
singularity at $t = 0$. The spacetime in the region $x + x_{0} \le -x_{5}$ can
be considered as representing the gravitational collapse of the perfect fluid.
To study the nature of the spacetime singularity at $x + x_{0} = - x_{5}$, let
us consider the quantity,
\bq
\lb{3a.5}
R_{,\alpha}R_{,\beta}g^{\alpha\beta} = \frac{4\alpha^{2}e^{2\left[(1-\alpha)x
- \alpha x_{0}\right]}}{9(- t)^{{2(\alpha - 1)}/{\alpha}}
\left[e^{-\alpha(x+x_{0})} - 1\right]^{2/3}} - 1,
\eq
from which we find that
\bq
\lb{3a.6}
\left(- t_{AH}\right)^{\frac{\alpha - 1}{\alpha}} = 
\frac{2\alpha e^{(1-\alpha)x - \alpha x_{0}}}{3\left[e^{-\alpha(x+x_{0})} -
1\right]^{1/3}},\; \left(R_{,\alpha}R_{,\beta}g^{\alpha\beta} = 0\right).
\eq

\vspace{1.cm}

\begin{figure}[htbp]
  \begin{center}
    \leavevmode
    \epsfig{file=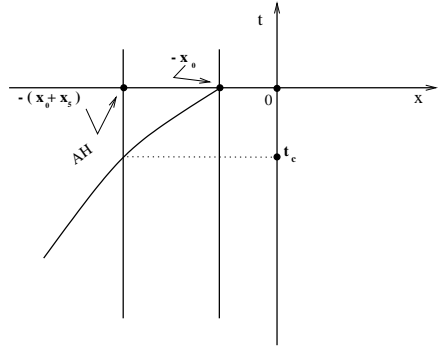,width=0.35\textwidth,angle=0}
    \caption{The spacetime described by the solutions of Eq.(\ref{3a.1}) for
$0 < \alpha < 1$ in the ($t, x$)-plane. It is singular on the
hypersurfaces $ a)\; t = 0,\; b)\; x = -  x_{0}$ and $ c)\; x  =- (x_{0} +
x_{5})$. At the moment $t = t_{c}$ the apparent horizon crosses the 
singular hypersurface $  x  = - (x_{0} + x_{5})$, and asymptotically approaches
to the point $(t, x ) = (0,  -  x_{0})$.}
    \label{fig3}
  \end{center}
\end{figure}

It can be shown that this hypersurface will cross the singular hypersurface $x
+ x_{0} = - x_{5}$ at the moment $t = t_{c}$, where $t_{c}$ is given by
\bq
\lb{3a.7}
\left(- t_{c}\right)^{\frac{1-\alpha}{\alpha}} = 
\frac{3}{2a} \left[e^{\alpha x_{5}} - 1\right]^{1/3}  
e^{(1-\alpha)x_{5} + x_{0}}. 
\eq
This can be seen clearly in the ($t, x$)-plane, as illustrated by Fig. 3. The
corresponding Penrose diagram is given by Fig. 4, from which we can see that
the spacetime singularity formed at $x + x_{0} = - x_{5}$ is  covered by the
apparent horizon at the beginning ($t < t_{c}$). As the fluid continues to
collapse, the apparent horizon starts to form after the formation of the
spacetime singularity, hence it becomes naked when   $t > t_{c}$.    

\vspace{1.cm}

\begin{figure}[htbp]
  \begin{center}
    \leavevmode
    \epsfig{file=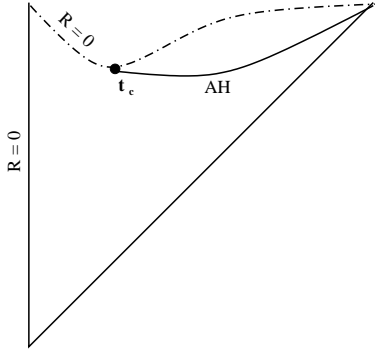,width=0.3\textwidth,angle=0}
    \caption{The Penrose diagram for the spacetime described by the solutions
of Eq.(\ref{3a.1}) for $0 < \alpha < 1$. The singularity
formed on the hypersurface  $x + x_{0} = - x_{5}$ is covered by the apparent
horizon when $t < t_{c}$ and becomes naked  when $t >  t_{c}$, where
$t_{c}$ is defined by Eq.(\ref{3a.7}).}
    \label{fig4}
  \end{center}
\end{figure}
{\bf b) Case $\; \alpha = 1$}: In this subcase, it can be shown that the
apparent horizon is given by
\bq
\lb{3a.8}
x + x_{0} = - x_{6},\; \left(R_{,\alpha}R_{,\beta}g^{\alpha\beta} = 0\right),
\eq
where
\bq
\lb{3a.9}
x_{6} \equiv \ln\left\{1 + \frac{8}{27}e^{-3x_{0}}\right\}.
\eq
Since
\bq
\lb{3a.10}
x_{6} - x_{5} = \ln\left(\frac{27 + 8 e^{-3x_{0}}}{81}\right) =
\cases{> 0, & $x_{0} < p^{*}$,\cr
= 0, & $x_{0} = p^{*}$,\cr
< 0, & $x_{0} > p^{*}$,\cr}
\eq
where $p^{*} \equiv -[\ln(27/4)]/3$, we find that, when $x_{0} < p^{*}$,
the apparent horizon always forms before the formation of the spacetime
singularity at $x + x_{0} = - x_{5}$, that is, the collapse now always forms
black holes. When $x_{0} = p^{*}$, the apparent horizon and the spacetime
singularity are formed on the same hypersurface, i.e., now the singularity is
marginally naked. When $x_{0} > p^{*}$, the apparent horizon always forms
after  the formation of the spacetime singularity, or in other words, now
the collapse always forms naked singularities.  
The contribution of the collapsing fluid to the total mass of such formed
black holes is given by  
\bq
\lb{3a.11}
M_{BH}(t) = \frac{2e^{-2x_{0}}}{9}r_{AH}(t),
\eq
where $r_{AH}(t)$ is a solution of Eq.(\ref{3a.8}). Thus, as $r_{AH}(t)
\rightarrow + \infty$, we find $M_{BH} \rightarrow + \infty$. Similar to the
case discussed in the last section, to obtain a black hole with finite mass,
we can cut the spacetime along the hypersurface $r = r_{0} = Const.$ and then
join the region $r \le r_{0}$ with an asymptotically flat region. By this way,
the resulting model will represent gravitational collapse of a ball with its
comoving radius $r_{0}$. At the moment $t = t_{f}$, where $t_{f}$ is a solution
of the
equation $r_{AH}(t_{f}) = r_{0}$, the  ball collapses completely inside the
horizon, and its contribution to the total mass of such formed black holes is
given by
\bq
\lb{3a.11a}
M_{BH} = \frac{2e^{-2x_{0}}}{9}r_{0},
\eq
which is always finite and non-zero for any given non-zero $r_{0}$.
It is interesting to note that in the present case the solutions may represent
critical phenomena. To have a definite answer to
this problem, we need to study the spectrum of perturbations of the
``critical" solution and show that it has only one unstable mode. This is
currently under our investigation.  

It is  very interesting to note that in
this case the black holes start to form with a mass gap, although    the
Einstein field equations are of scale invariance, and the spacetime has
self-similarity of the first kind  \cite{Gu00,Wang01}.  Thus, the solutions
studied in the present case show clearly that even the  solutions have 
homothetic self-similarity, the formation of black holes does not necessarily
start with an infinitesimal mass.  

\vspace{1.cm}
 
\begin{figure}[htbp]
  \begin{center}
    \leavevmode
    \epsfig{file=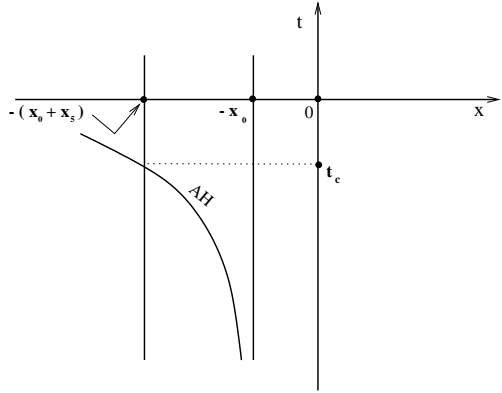,width=0.4\textwidth,angle=0}
    \caption{ The spacetime described by the solutions of Eq.(\ref{3a.1}) for
$1 < \alpha < 3/2$  in the ($t, x$)-plane. It is singular on the
hypersurfaces $ a)\; t = 0,\; b)\; x = -  x_{0}$ and $ c)\; x  =- (x_{0} +
x_{5})$. At the moment $t = t_{c}$ the apparent horizon crosses the 
singular hypersurface $  x  = - (x_{0} + x_{5})$, and asymptotically approaches
to the one $x = -  x_{0}$.}
    \label{fig5}
  \end{center}
\end{figure}

{\bf c) Case $\; 1 < \alpha < \frac{3}{2}$}: In this subcase, it can be shown
that the spacetime is also singular on the hypersurfaces given by
Eq.(\ref{3a.3}) and the apparent horizon  is given by Eq.(\ref{3a.6}). 
In the ($t, x$)-plane, it is given by Fig. 5, from which we can see that
it  also crosses the singular hypersurface $x + x_{0} = - x_{5}$ once, but
in contrast to the subcase $0 < \alpha < 1$, now the singularity initially is
naked and becomes covered by the apparent horizon after the moment $t =
t_{c}$, as shown by Fig. 6.

{\bf d) Case $\;   \alpha = \frac{3}{2}$}: From Eq.(\ref{3a.2}) we find that 
$\rho = 0 = p$. That is, in this subcase the spacetime is vacuum. The metric
coefficients are given by
\bqn
\lb{3a.13}
y(x) &=& \left[e^{3(x+x_{0})/2} - 1\right]^{-1},\nb\\
S(x) &=& e^{-(x+x_{0})}
\left[e^{3(x+x_{0})/2} - 1\right]^{2/3}.
\eqn
Defining a new radial coordinate $\tilde{r}$ via the relations,
\bq
\lb{3a.14}
\tilde{r}\equiv \frac{2}{3}e^{3x_{0}/2}r^{3/2},
\eq
we find that the metric can be written as
\bq
\lb{3a.15}
ds^{2} = d\tilde{\tau}^{2} -
r^{2/3}_{g}\left\{\frac{d\tilde{r}^{2}}{\left[\frac{3}{2}(\tilde{r}-\tilde{\tau}
)\right]^{2/3}} - \left[\frac{3}{2}(\tilde{r}-  \tilde{\tau})\right]^{4/3}
d^{2}\Omega\right\}, 
\eq 
where
\bq
\lb{3a.16}
\tilde{\tau} = -t, \;\;\; r_{g} = e^{-3x_{0}}.
\eq
This is exactly the Schwarzschild solution written in the Lemaitre coordinates 
\cite{LL75}, with $r_{g}$ being the Schwarzschild radius. From this case we
 can see that the parameter $x_{0}$ is related to the total mass of the
Schwarzschild black hole.

\vspace{1.cm}

\begin{figure}[htbp]
  \begin{center}
    \leavevmode
    \epsfig{file=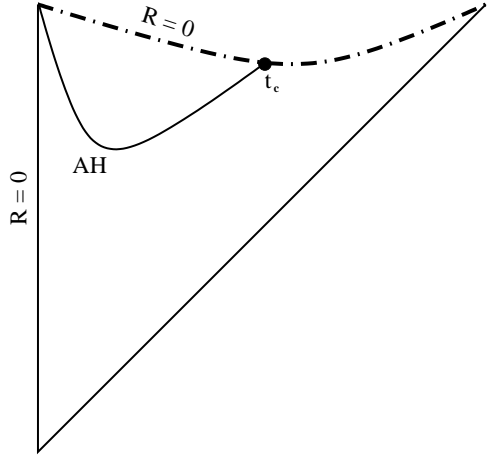,width=0.39\textwidth,angle=0}
\caption{The Penrose diagram for the spacetime described by the solutions
of Eq.(\ref{3a.1}) for $1 < \alpha < 3/2$. The singularity formed on the
hypersurface  $x + x_{0} = - x_{5}$ is naked for
$t < t_{c}$ and covered by the apparent horizon when $t >  t_{c}$, where
$t_{c}$ is defined by Eq.(\ref{3a.7}).}
    \label{fig6}
  \end{center}
\end{figure}

{\bf e) Case $\; \alpha > \frac{3}{2}$}: In this subcase the energy density of
the fluid  takes the form
\bq
\lb{3a.17}
\rho = \frac{4(2\alpha - 3)}{9r^{2}_{1}}\left\{t^{2}\left[1 -
e^{\alpha(x+x_{0})}\right]
\left[e^{\alpha(x+x_{0})} + \frac{2\alpha -
3}{3}\right]\right\}^{-1}. 
\eq
Thus, $\rho \ge 0$ requires
\bq
\lb{3a.18}
x + x_{0} \le 0, \;\; (\rho \ge 0).
\eq

\vspace{1.cm}

\begin{figure}[htbp]
  \begin{center}
    \leavevmode
    \epsfig{file=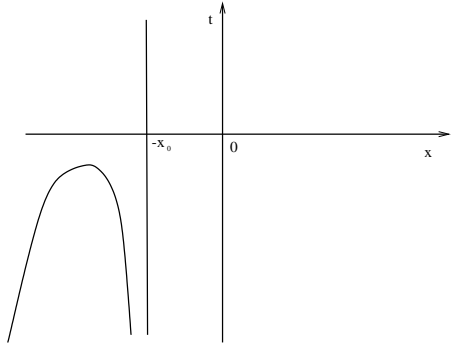,width=0.36\textwidth,angle=0}
    \caption{The spacetime described by the solutions of Eq.(\ref{3a.1}) for
$ \alpha > 3/2$ in the ($t, x$)-plane. It is singular on the
hypersurface $ x = -  x_{0}$ and the energy density of the fluid is
non-negative only in the region $ x   \le - x_{0}$.  }
    \label{fig7}
  \end{center}
\end{figure}

\vspace{1.cm}

\begin{figure}[htbp]
  \begin{center}
    \leavevmode
    \epsfig{file=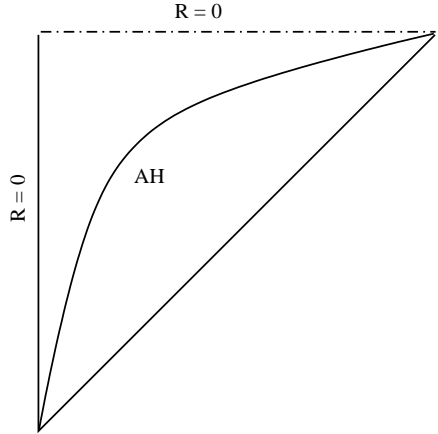,width=0.35\textwidth,angle=0}
   \caption{The Penrose diagram for the spacetime described by the solutions
of Eq.(\ref{3a.1}) for $\alpha > 3/2$. The apparent horizon now
always forms before the formation of the spacetime singularity at $x +
x_{0} = 0$.  }
    \label{fig8}
  \end{center}
\end{figure}

\noindent In this region the spacetime  is singular only on the hypersurface
$
 x + x_{0} = 0.
$
On the other hand, the apparent horizon in the present
case is still given by Eq.(\ref{3a.6}). In the ($t, x$)-plane, this
hypersurface 
is shown by Fig. 7. The corresponding Penrose diagram is given by
Fig. 8,  from which we can see that now the apparent horizon always forms
before the formation of the  spacetime singularity, that is, the solutions
now represent the formation of black holes. 

The contribution of the fluid to the total mass of such formed black
holes is given by 
\bq
\lb{3a.19}
M_{BH}(t) = \frac{r_{AH}(t)}{2}\left[e^{-\alpha(x_{AH} + x_{0})} - 1
\right]^{2/3},
\eq
where $r_{AH}$ and $x_{AH}$ are the solution of Eq.(\ref{3a.6}). When the
spacetime is first cut along the hypersurface $r= r_{0}$ and then joined with
an asymptotically flat region, and the contribution of the fluid to the total
mass of such formed black holes is given by 
\bq
\lb{3a.19a}
M_{BH} = \frac{r_{0}}{2}\left[e^{-\alpha(x_{f} + x_{0})} - 1
\right]^{2/3},
\eq
where $x_{f} = \ln[r_{0}/(- t_{f})^{1/\alpha}]$, and $t_{f}$ denotes the
moment when the ball collapses completely inside the apparent horizon, given
by $r_{0} = r_{AH}(t_{f})$. One can show that this mass is also finite and
non-zero for any given non-zero $r_{0}$.

%%%%%%%%%%%%%%%%%%%%%%%%%%%%%%%%%%%%%%%%%%%%%%%%%%%%%%%%%%%%%%%%%%%%%%%%%%

\subsection{Case $0 < p_{0} < \frac{1}{3}$}

In this case, it can be shown that the corresponding solutions are given by
\bqn
\lb{3a.20}
y(x)&=& \frac{1}{3}\left\{\beta\tanh\left[\frac{\beta}{2} (x + x_{0})\right] 
- \alpha\right\},\nb\\ 
S(x) &=& e^{-\alpha x/3}\cosh^{2/3}\left[\frac{\beta}{2}(x +
x _{0})\right], 
\eqn
where  $\beta \equiv \alpha|1 -
3p_{0}|^{1/2}$. Then, the energy density of the fluid is given by
\bqn
\nb
\rho & = & \frac{\beta(2\alpha - 3)\left\{(1- 3p_{0})^{-1/2} - 
\tanh\left[\frac{\beta}{2}(x + x _{0})\right]\right\}}{3\alpha^{2}r^{2}_{1}t^{2}
\left\{\tanh\left[\frac{\beta}{2}(x + x _{0})\right] + A\right\}} \\
 \lb{3a.21} & & \; \; \times
\left\{\tanh\left[\frac{\beta}{2}(x + x _{0})\right] + B\right\},
\eqn
where
\bq
\lb{3a.22}
A  \equiv  \frac{3 - \alpha}{\alpha\left|1- 3p_{0}\right|^{1/2}},\;\;\;
B  \equiv  \frac{3 \alpha p_{0} - (2\alpha - 3)}{\left|1-
3p_{0}\right|^{1/2}(2\alpha - 3)}.
\eq
From the above equation it can be shown that, when $0 < \alpha < \alpha_{1}$,
we have $A > 1$; when  $\alpha_{1} \le \alpha \le \alpha_{2}$, we have $ -1 \le
A \le + 1$; and when $\alpha > \alpha_{2}$, we have $A < -1$, where
\bq
\lb{3a.23}
\alpha_{1}  \equiv  \frac{3}{1 + (1 - 3p_{0})^{1/2}},\;\;\;
\alpha_{2}  \equiv  \frac{3}{1 - (1 - 3p_{0})^{1/2}}.
\eq
When $0 < \alpha \le 3/2$, we have $B < - (1 - 3p_{0})^{-1/2} < -1$; when 
$3/2 <  \alpha < \alpha_{1}$, we have $   B > + 1$; when $\alpha_{1} \le
\alpha \le \alpha_{2}$, we have $- 1 \le B \le + 1$; and when $\alpha >
\alpha_{2}$, we have $B < -1$ [cf. Fig. 9]. Thus, from Eq.(\ref{3a.21}) we
find that $\rho$ is always non-negative at any given point of the 
spacetime when $0 < \alpha < \alpha_{1}$ or when $ \alpha > \alpha_{2}$. It is 
 singular on the hypersurface $t = 0$. This hypersurface divides the
spacetime into two disconnected regions, $ t \ge 0$ and $t \le 0$. In the
region $t \ge 0$, the spacetime can be considered as representing a
cosmological model with its initial singularity at $t = 0$. In the region $ t
\le 0$, the spacetime can be interpreted as representing the gravitational
collapse of the perfect fluid. In this region, an apparent horizon is 
formed on the hypersurface, given by 
\bq
\lb{3a.24}
 \left(- t_{AH}\right)^{\frac{\alpha - 1}{\alpha}}   = 
\frac{\alpha e^{(3-\alpha)x/3} }{3 \cosh^{1/3}\left[\frac{\beta}{2}(x +  
x_{0})\right]}  \times
\left\{\cosh\left[\frac{\beta}{2}(x + x
_{0})\right] - (1 - 3p_{0})^{1/2}\right\}, 
\left(R_{,\alpha}R_{,\beta}g^{\alpha\beta} = 0\right). 
\eq

\vspace{1.cm}

\begin{figure}[htbp]
  \begin{center}
    \leavevmode
    \epsfig{file=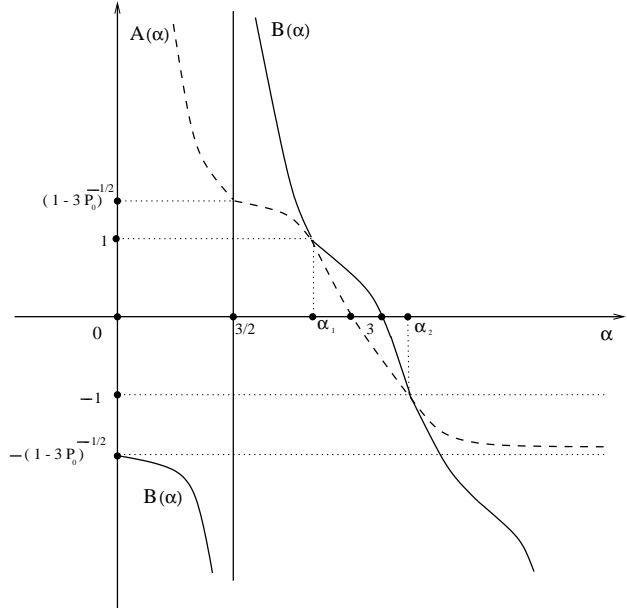,width=0.5\textwidth,angle=0}
    \caption{The curves of the functions $A(\alpha)$ and $B(\alpha)$,
defined by Eq.(\ref{3a.22})  versus $\alpha$.}
    \label{fig9}
  \end{center}
\end{figure}

It is not difficult to see that this hypersurface is formed always before the
formation of the spacetime singularity at $t = 0$. Thus in the present case
the collapse always forms black holes, and the contribution of the fluid to the
black hole mass   is given by 
\bq
\lb{3a.25}
M_{BH}(t) = \frac{r_{AH}(t)}{2}\cosh^{2/3}\left[\frac{3}{2}(x_{AH} +
x_{0})\right]e^{-\alpha x_{AH}/3},
\eq
where $r_{AH}(t)$ and $x_{AH}$ are the solutions of Eq.(\ref{3a.24}). Similar
to the cases discussed above, the mass of such formed black holes now becomes
also infinitely large as $r_{AH}(t) \rightarrow + \infty$. Thus, in this case
we also need to cut the spacetime along the hypersurface $r = r_{0}$ and then
join the region $ r \le r_{0}$ to an asymptotically flat region. Once this
is done, it is not difficult to see that  
\bq
\lb{3a.25a}
M_{BH} = \frac{r_{0}}{2}\cosh^{2/3}\left[\frac{3}{2}(x_{f} +
x_{0})\right]e^{-\alpha x_{f}/3},
\eq
where $x_{f}$ is given by $x_{f} = \ln [r_{0}(-t_{f})^{-1/\alpha}]$, and
$t_{f}$ denotes the moment when the ball of perfect fluid collapses
completely inside the apparent horizon, which is given by Eq.(\ref{3a.24})
with $ r_{AH}(t_{f}) = r_{0}$. Clearly, for any given non-zero $r_{0}$,
$M_{BH}$ is non-zero. That is, in the present case the black holes start to
form with a mass gap, too.

\vspace{1.cm}

\begin{figure}[htbp]
  \begin{center}
    \leavevmode
    \epsfig{file=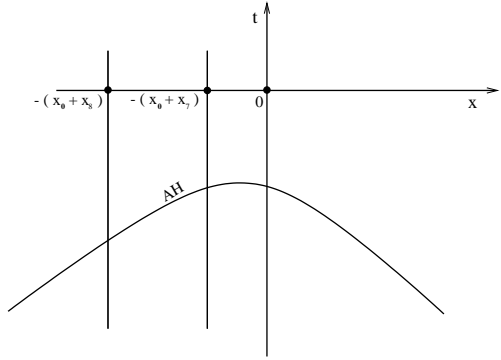,width=0.4\textwidth,angle=0}
   \caption{The spacetime described by the solutions of Eq.(\ref{3a.20}) for
$ \alpha_{1} \le \alpha \le \alpha_{2}$ in the ($t, x$)-plane.  $\rho$ is
non-negative only in the region $ x \ge - (x_{0} + x_{7})$ or in the region $
x  \le - ( x_{0}+ x_{8})$. It is singular on the hypersurface $x = - (x_{0} +
x_{7})$.}
    \label{fig10}
  \end{center}
\end{figure}

When $ \alpha_{1} \le \alpha \le \alpha_{2}$, $\rho$ is non-negative only in
the region $ x + x_{0} \ge - x_{7}$ or in the region $ x + x_{0} \le - x_{8}$,
where \bq
\lb{3a.26}
x_{7} \equiv \frac{2}{\beta}\tanh^{-1}(A),\;\;\;\;
x_{8} \equiv \frac{2}{\beta}\tanh^{-1}(B).
\eq
Since $ B \ge A$ for $\alpha_{1} \le \alpha \le \alpha_{2}$, we find $ x_{8}
\ge x_{7}$, where equality holds only when $\alpha = \alpha_{1}$, or $\alpha =
\alpha_{2}$. The spacetime is singular on the hypersurface $t = 0$ and $x +
x_{0} = - x_{7}$. Once again, the region $ t \ge 0$ can be considered as
representing a cosmological model. In the region $ x + x_{0} \le - x_{7}$, on
the other hand, we find that
\bq
\lb{3a.27}
\rho = \cases{ - \infty, & $x + x_{0} = - x_{7}$,\cr
< 0, & $- x_{8} < x + x_{0} < - x_{7}$,\cr
= 0, & $ x+ x_{0} = - x_{8}$,\cr
> 0, & $ x+ x_{0} < - x_{8}$.\cr}
\eq
In this region, the apparent horizon is still  given by Eq.(\ref{3a.24}), from
which we find that, as $x \rightarrow \pm \infty$, we have 
$(- t)^{(\alpha - 1)/\alpha} \rightarrow + \infty$.  
In the ($t, x$)-plane, it is given by a curve that crosses both the
hypersurfaces $x + x_{0} = - x_{7}$ and $x + x_{0} = - x_{8}$ [cf. Fig.    
10]. The corresponding Penrose diagram is given by Fig. 11, the physics of
which is unclear.  

\vspace{1.cm}

\begin{figure}[htbp]
  \begin{center}
    \leavevmode
    \epsfig{file=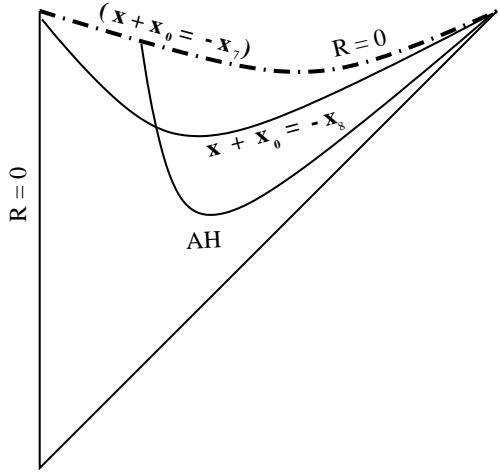,width=0.4\textwidth,angle=0}
    \caption{The Penrose diagram for the spacetime described by the solutions
of Eq.(\ref{3a.20}) for $ \alpha_{1} \le \alpha \le \alpha_{2}$. In the region
$ - x_{7} \le x + x_{0} \le - x_{8}$, the energy density of the fluid becomes
negative.}
    \label{fig11}
  \end{center}
\end{figure}

%%%%%%%%%%%%%%%%%%%%%%%%%%%%%%%%%%%%%%%%%%%%%%%%%%%%%%%%%%%%%%%%%%%%%%%%%%

\subsection{Case $ p_{0} = \frac{1}{3}$}

In this case, it can be shown that Eqs.(\ref{3.2}) and (\ref{3.3}) have the
following solutions,
\bqn
\lb{3.4}
y &=& \frac{2}{3(x+ x_{0})} - \frac{\alpha}{3},\;\;\;\;
S = (x + x_{0})^{2/3} e^{- \alpha x/3},\nb\\
\rho &=& \frac{1}{3\alpha^{2}r^{2}_{1}t^{2}(x + x_{0})\left[(3 - \alpha)(x +
x_{0}) + 2\right]}\nb\\
& & \times \left\{\alpha^{2}(3 - \alpha)(x + x_{0})^{2} \right.\nb\\
& & \left. - 6\alpha(2 - \alpha)(x + x_{0}) + 4(3 - 2\alpha)\right\},\nb\\
p &=& \frac{1}{3r^{2}_{1}t^{2}}.
\eqn

{\bf A) Case $\; 0 <  \alpha < 1$}: In this subcase from Eq.(\ref{3.4}) we can
see that the spacetime is singular on the hypersurfaces
\bq
\lb{3.21}
a)\; t = 0,\;\;\;\;\;\; 
b)\; x + x_{0} = 0,\;\;\;\;\;\;
c)\; x + x_{0} = - \frac{2}{3 - \alpha}.
\eq
It is not difficult to show that now the singularity at $x + x_{0} = - {2}/{(3
- \alpha)}$ is first formed and the one at $t =0$ is last formed. The spacetime
in the region $ t\ge 0$ may be considered as representing cosmological model
with its initial singularity of the spacetime at $t = 0$. The region $x + x_{0}
\le  - {2}/{(3 - \alpha)}$ can be considered as representing the gravitational
collapse of the perfect fluid. To study the nature of this singularity, let us
consider the formation of apparent horizons, given by
\bq
 \lb{3.22}
\left(- t_{AH}\right)^{\frac{2(1-\alpha)}{\alpha}} 
=  \frac{9(x+x_{0})^{2/3}}{\left[2 - \alpha
(x+x_{0})\right]^{2}}e^{-\frac{2(3-\alpha)}{3}x},\;
\left(R_{,\alpha}R_{,\beta}g^{\alpha\beta}  =  0\right). 
\eq
In the ($t, x$)-plane, this curve is similar to that given in Fig.  3,  if 
$x_{5}$ is replaced by $2/(3-\alpha)$. Thus, in this case the singularity 
at $x + x_{0}= - 2/(3-\alpha)$ is covered upto the moment $t = t_{c}$,
where $t_{c}$ now is given by
\bq
\lb{3.23}
\left(-t_{c}\right)^{\frac{(1-\alpha)}{\alpha}} =
\frac{1}{3}\left(\frac{3- \alpha}{2}\right)^{2/3}e^{2/3}.
\eq
After this moment, the singularity becomes naked. The corresponding Penrose diagram 
is given by Fig. 4.

{\bf B) Case $\; \alpha = 1$}:
As we mentioned previously, when $\alpha = 1$, the corresponding solutions have
the  self-similarity of the first kind. Setting $\alpha = 1$ in the above
expressions, we find that
\bq
\lb{3.5}
\rho = \frac{(x + x_{0})^{2} - 3(x + x_{0}) + 2}{3r^{2}_{1}t^{2}(x +
x_{0})\left[(x + x_{0}) + 1\right]},\;\;\;\;\;
 p = \frac{1}{3r^{2}_{1}t^{2}}.
\eq
Clearly, the spacetime is also singular on the hypersurfaces given by
Eq.(\ref{3.21}).  Since now we have $e^{x} = r/(-t)$, 
we can see that these singular hypersurfaces are straight lines in the ($t,
r$)-plane. The spacetime in the region $t \ge 0$ may be
interpreted as representing cosmological model with an initial spacetime
singularity at $ t = 0$. The region $x + x_{0} \le - 1$ may be considered as
representing the gravitational collapse of the perfect fluid starting at $t =
- \infty$. To study the nature of the singularity located on the hypersurface
$x + x_{0} = -1$, let us, following the analysis given above,
first consider the formation of apparent horizon in this region,
\bq
\lb{3.8}
R_{,\alpha} R_{,\beta} g^{\alpha\beta} = \frac{e^{4x/3}}{9(x +
x_{0})^{2/3}} \left\{Y_{1}(x) - Y_{2}(x)\right\},
\eq
where
\bq
\lb{3.9}
Y_{1}(x) \equiv \left[2 - (x + x_{0})\right]^{2},\;\;\;
Y_{2}(x) \equiv 9(x + x_{0})^{2/3}e^{-4x/3}.
\eq
It is easy to show that $Y_{2}(x)$ has one minimal at $x + x_{0} = 0$ and one
maximal at $x + x_{0} = 1/2$. When $x \rightarrow - \infty$ it diverges
exponentially, and when $x \rightarrow + \infty$ it goes to zero
exponentially. On the other hand, $Y_{1}(x)$ is a parabola with its minimum
located at $x + x_{0} = 2$ [cf. Fig. 12]. Thus, in general the equation
$Y_{1}(x) = Y_{2}(x)$ has three real  roots, say, $x_{9}, \; x_{10}$ and $
x_{11}$. Without loss of generality, we assume that $x_{11} > x_{10} > x_{9}$.
Then, we can see that $x = x_{9}$ represents the
outmost trapped surface, i.e., the apparent horizon. Introducing a new
parameter $D$ via the relation,
\bq
\lb{3.10}
D \equiv - \left(x_{9} + x_{0} + 1\right),
\eq
we find that the hypersurface $x = x_{9}$ can be written as 
\bq
\lb{3.11}
x + x_{0} = - ( 1 + D), 
\; \left(R_{,\alpha}R_{,\beta}g^{\alpha\beta}
= 0\right).
\eq
Thus, when $D > 0$, the apparent horizon always forms before the formation of
the spacetime singularity at $x + x_{0} = - 1$. That is, in this case the
gravitational collapse of the perfect fluid always forms black holes. When $D
< 0$, the apparent horizon always forms after the formation of the spacetime
singularity at $x + x_{0} = - 1$, namely, now the collapse always
forms naked singularities. When $D = 0$, the apparent horizon and the
spacetime  singularity are formed on the same hypersurface $x + x_{0} = - 1$,
and now the singularity is marginally naked.  In
the last case, it can be shown that 
\bq
\lb{3.12}
x_{9} = x_{0} = 0,\; (D =0),
\eq
and the corresponding solution is given by
\bq
\lb{3.13}
S = x^{2/3}e^{-x/3},\;\;\; 
y = \frac{2 - x}{3x}, \;  (D =0).
\eq
Similar to the case $p_{0} =0$ and $\alpha =1$ discussed above in this
section, these solution may also represent critical collapse. To have a
definite answer to this problem, we need to study perturbations of the
``critical" solution, which is out of the scope of this paper, and we hope to
return to this problem in another occasion.

On the other hand, the mass function defined by
Eq.(\ref{2.10}) now takes the form, \bq \lb{3.14}
m(t,r) = \frac{rY_{1}(x)}{18} e^{x}.
\eq
Thus, on the apparent horizon $x = x_{9}$, we have
\bq
\lb{3.15}
M_{BH}(t) = \frac{(3+D)^{2}e^{x_{9}}}{18}r_{AH}(t),
\eq
which shows that, as $r_{AH}(t) \rightarrow + \infty$, the total mass of black
hole
becomes infinitely large. Similar to the cases considered above, to have a
black hole with finite mass, we can make a ``surgery" to the spacetime. By this
way, we can see that the resultant solution will represent a collapsing ball
with a finite radius $r_{0}$, and its contribution to the total mass of black
hole is given by 
\bq
\lb{3.16}
M_{BH} = \frac{(3+D)^{2}e^{x_{9}}}{18}r_{0},\; (D \ge 0),
\eq
which is finite and non-zero for any given non-zero $r_{0}$.

{\bf C) Case $\; 1 <  \alpha < 3$}: In this case the spacetime singularities
and apparent horizon are still given by Eqs.(\ref{3.21}) and (\ref{3.22}),
respectively. It can be shown that the curve that represent the apparent horizon
in the ($t, x$)-plane now is similar to that given in Fig. 5, that is, in
the present case the singularity at $x + x_{0} = - 2/(3 - \alpha)$ is initially
naked. As the fluid is collapsing until the moment 
$t_{c}$ given by Eq.(\ref{3.23}), the apparent horizon starts to form. The 
corresponding Penrose diagram is given by Fig. 6.  

\vspace{1.cm}

\begin{figure}[htbp]
  \begin{center}
    \leavevmode
    \epsfig{file=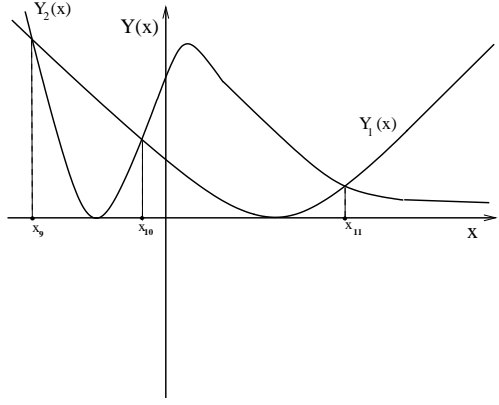,width=0.4\textwidth,angle=0}
   \caption{The curves of the functions $Y_{1}(x)$ and $Y_{2}(x)$,
defined by Eq.(\ref{3.9})   versus $x$. The equation
$Y_{1}(x) = Y_{2}(x)$  in general has three real roots, $x_{9}, \; x_{10}$ and $
x_{11}$, with $x_{11} > x_{10} > x_{9}$.}
    \label{fig12}
  \end{center}
\end{figure}
 
{\bf D) Case $\; \alpha = 3$}: In this subcase, the corresponding physical
quantities are given by \bqn
\lb{3.17}
\rho &=& \frac{3(x + x_{0}) - 2}{9r^{2}_{1}t^{2}(x + x_{0})},\nb\\
R_{,\alpha}R_{,\beta}g^{\alpha\beta} & =& 
\frac{\left[3(x + x_{0}) - 2\right]^{2}}{9(-3t)^{4/3}(x + x_{0})^{2/3}} - 1,
\eqn
which show that the spacetime is singular at
\bq
\lb{3.18}
 a)\; t = 0,\;\;\;\;\;\; 
b)\; x + x_{0} = 0.
\eq
The location of the apparent horizon is given by
\bq
\lb{3.19}
\left(- t_{AH}\right)^{4/3} = \frac{\left[(x + x_{0}) -
\frac{2}{3}\right]^{2}}{(x + x_{0})^{2/3}},\; 
\left(R_{,\alpha}R_{,\beta}g^{\alpha\beta} = 0\right),
\eq
from which we can see that in the ($t, x$)-plane it is given by the curved 
given in Fig. 7. Consequently, the solutions in this case represent
gravitational collapse that always forms black holes.  

{\bf E) Case $\; \alpha > 3$}: In this case, following the same routine given
above, it is not difficult to show that the spacetime singularity at $x +
x_{0} = 0$, formed due to the collapse of the perfect fluid, is also covered by
an apparent horizon. It can be shown that in the last two subcases the mass of such
formed black holes is always finite and non-zero. 

\subsection{Case $ p_{0} > \frac{1}{3}$}

In this case,   the corresponding solutions are given by
\bqn
\lb{3.22a}
y(x)&=& -\frac{1}{3}\left\{\beta\tan\left[\frac{\beta}{2} (x + x_{0})\right] 
+ \alpha\right\},\nb\\ 
S(x) &=& e^{-\alpha x/3}\cos^{2/3}\left[\frac{\beta}{2}(x +
x _{0})\right], 
\eqn
where      $\beta$ is given as that 
in Eq.(\ref{3a.20}).   Then, the energy density of
the fluid is given by  
\bqn
\nb 
\rho & = & \frac{\beta(2\alpha - 3)\left\{(3p_{0} - 1)^{-1/2} + 
\tan\left[\frac{\beta}{2}(x + x _{0})\right]\right\}}{3\alpha^{2}r^{2}_{1}t^{2}
\left\{\tan\left[\frac{\beta}{2}(x + x _{0})\right] - A\right\}}
\\ \lb{3.23a} & & \; \; \times
\left\{\tan\left[\frac{\beta}{2}(x + x _{0})\right] - B\right\},
\eqn
where $A$ and $B$ are given by Eq.(\ref{3a.22}). From the above expression we
can see that $\rho$ is non-negative only in certain regions and the spacetime
is singular on various hypersurfaces. The solutions in this case
cannot be interpreted as representing gravitational collapse of perfect fluid.

\section{Concluding Remarks}  

In Sec. III, we have studied the self-similar solutions of the zeroth kind, and
found that some represent cosmological models and some represent gravitational
collapse, while the others have no physical meanings. The ones that
represent gravitational collapse are given by $p = 0$, i.e., dust fluid. These
dust fluid solutions always collapse to form black holes with finite
and non-zero mass.

In Sec. IV, self-similar solutions of both the first and the second kinds
have been studied. In particular, it has been found that the self-similar
solutions of the first kind ($\alpha = 1$) with $p_{0} = 0 $ or $p_{0} = 1/3$
may represent critical collapse but in the sense that now the ``critical"
solution separates solutions that form black holes to the solutions that form
naked singularities.  In this case the formation of black holes also starts
with a mass gap. To show explicitly that these solutions indeed represent
critical collapse, the analysis of spectrum of perturbations of these
``critical" solutions is needed, which are currently under our investigation.
The solutions with $ p_{0} = 0$ and $\alpha > 3/2$, the ones with $0 < p_{0} <
1/3$ and $ \alpha _{1} < \alpha < \alpha_{2}$, and the ones with $p_{0} = 1/3$
and $\alpha \ge 3$  also represent gravitational collapse, and the collapse
always forms black holes with finite and non-zero mass.  The solutions with
$p_{0} = 0$ and $0 < \alpha < 1$ and the ones with $p_{0} = 1/3$ and 
$0 < \alpha < 1$ represent the formations of spacetime
singularities that are covered by apparent horizons at the beginning of the
collapse and late become naked, while the ones with  $p_{0} = 0$ and $1 <
\alpha < 3/2$ and the ones with $p_{0} = 1/3$ and $1 < \alpha < 3$ represent 
the formations of spacetime singularities that are
naked  at the beginning of the collapse and late
become covered by apparent horizons. All the rest of the solutions
can be either considered as representing cosmological models with an 
initial spacetime singularity or have no physical meanings. 

In review of all the above, one can see that the BC solutions with
self-similarity of the zeroth and second kinds seem irrelevant to critical
phenomena in gravitational collapse, and the only possible candidates for
critical collapse are those solutions with self-similarity of the first
(homothetic) kind with $p_{0} = 0$ or $p_{0} = 1/3$, given in Sec.IV.

\section*{ACKNOWLEDGMENTS}

One of the authors (AZW) would like to express his gratitude to Professor W.M.
Suen and his group for valuable suggestions and discussions on Critical
Collapse.  Part of the work was done when he was  visiting the
McDonnell Center for the Space Sciences, Department of Physics, Washington
University, St. Louis, USA. He would like to thank  the Center for
hospitality. The financial  assistance from the Center and FAPERJ (AZW),
which made this visit possible, is gratefully acknowledged. We would also like
to express our gratitude to P.M. Benoit for sending us her unpublished Ph.D.
thesis. Finally we thank CNPq (JFVR, AZW) and UERJ (AZW) for  financial 
assistance.

\section*{Appendix}

\renewcommand{\theequation}{A.\arabic{equation}}
\setcounter{equation}{0}

The metric for spacetimes with spherical symmetry can be cost in the general
form, \bq
\lb{A.1}
ds^2 = r^{2}_{1}\left\{e^{2 \Phi(t,r) }dt^2 - e^{2 \Psi(t,r)}dr^2 - r^2
S(t,r)^2 d\Omega^2\right\},
 \eq
where $d\Omega^{2} \equiv d\theta^2 + \sin(\theta)^2d\varphi^{2}$, and $r_{1}$
is a constant and has  dimension of length, $l$. Then, it is easy to show that the
coordinates $\{x^{\mu}\} = \{t, \; r, \; \theta, \; \varphi\}$, the
Christoffel symbols, $\Gamma^{\mu}_{\lambda\nu}$, the Riemann tensor,
$R^{\sigma}_{\mu\nu\lambda}$, the Ricci tensor, $R_{\mu\nu}$, and the Einstein
tensor, $G_{\mu\nu}$, are all {\em dimensionless}, while the Ricci scalar, $R$,
has the dimension of $l^{-2}$, and the Kretschmann scalar, $I \equiv 
R^{\sigma\mu\nu\lambda} R_{\sigma\mu\nu\lambda}$, has the dimension of
$l^{-4}$.  

For the metric (\ref{A.1}), we find that the non-vanishing 
Christoffel symbols  are given by 
\bqn
\lb{A.2} 
\Gamma^{0}_{00} &=& \Phi_{,t},\;\;\; \Gamma^{0}_{01} =
\Phi_{,r},\;\;\; \Gamma^{0}_{11} = e^{2(\Psi -\Phi)}\Psi_{,t},\nb\\
\Gamma^{0}_{22} &=& r^{2}S  e^{ -2\Phi}S_{,t},\;\;\;\;
\Gamma^{0}_{33} = r^{2}S\sin^{2}\theta  e^{ -2\Phi}S_{,t},\nb\\
\Gamma^{1}_{00} &=& e^{2(\Phi -\Psi)}\Phi_{,r},\;\;\;\;
\Gamma^{1}_{01} = \Psi_{,t},\;\;\;
\Gamma^{1}_{11} = \Psi_{,r},\nb\\
\Gamma^{1}_{22} &=& - rS  e^{ -2\Psi}\left(rS_{,r} + S\right),\;\;\;
\Gamma^{1}_{33} = - rS \sin^{2}\theta e^{ -2\Psi}\left(rS_{,r} + S\right),\nb\\
\Gamma^{2}_{02} &=& \frac{S_{,t}}{S},\;\;\;
\Gamma^{2}_{12} = \frac{ rS_{,r} + S}{rS},\;\;\;
\Gamma^{2}_{33} = -\sin\theta \cos\theta,\nb\\
\Gamma^{3}_{03} &=& \frac{S_{,t}}{S},\;\;\;
\Gamma^{3}_{13} = \frac{ rS_{,r} + S}{rS},\;\;\;
\Gamma^{3}_{23} =  \frac{\cos\theta}{\sin\theta}, 
\eqn
while the non-vanishing  components of the  Einstein tensor are given by
\bqn
\lb{A.3a}
G_{tt} &=& - \frac{e^{-2\Psi}}{r^2 S^2} \left\{ 
e^{2\Phi}\left[2r^{2}SS_{,rr} + rS_{,r}\left(rS_{,r} + 6S\right) -
2rS\left(rS_{,r} + S\right)\Psi_{,r} + S^{2} - 
e^{2\Psi}\right]\right.\nb\\
& & \left. - r^{2}e^{2\Psi}S_{,t}\left(2S\Psi_{,t} + S_{,t}\right)\right\},\\
\lb{A.3b}
G_{tr} &=& -\frac{2}{rS} \left[r S_{,tr} - \left(rS_{,r} + S\right)\Psi_{,t}
- S_{,t}\left(r\Phi_{,r} - 1\right)\right],\\
\lb{A.3c}
G_{rr} &=&  \frac{e^{-2\Phi}}{r^2 S^2} \left\{ 
e^{2\Phi}\left[2rS\left(rS_{,r} + S\right)\Phi_{,r} +
rS_{,r}\left(rS_{,r} + 2S\right) + S^{2} - e^{2\Psi}\right]\right.\nb\\
& & \left. - r^{2} e^{2\Psi}\left[2SS_{,tt} + S_{,t}\left(S_{,t} -
2S\Phi_{,t}\right)\right]\right\},\\
\lb{A.3d}
G_{\theta \theta}  &=&  rS e^{-2(\Phi + \Psi)}\left\{ 
e^{2\Phi}\left[r\left(S\Phi_{,rr} + S_{,rr}\right) +
rS\Phi_{,r}\left(\Phi_{,r} - \Psi_{,r}\right)\right.\right. \nb\\
& & \left. \;\;\;\;\;\; \;\;\;\;\;\;\;\;\;\;\;\;\;\;\;\;\;\;\;\;
 + \left(rS_{,r} + S\right)\left(\Phi_{,r} - \Psi_{,r}\right) +
2S_{,r}\right]\nb\\ 
& & \left. - re^{2\Psi}\left[S\Psi_{,tt} + S_{,tt} -
\left(S\Psi_{,t} + S_{,t}\right)\left(\Phi_{,t} -
\Psi_{,t}\right)\right]\right\},
\eqn
where $(\;)_{,\alpha} \equiv \partial(\;)/\partial x^{\alpha}$, ect.

\subsection{Solutions with Self-Similarity of the zeroth Kind}

To study solutions with kinematic self-similarity of the zeroth kind,  let us
introduce two new dimensionless variables, $x$ and $\tau$, via the relations
\bq
\lb{B.1}
x = \ln(\xi) =  - t + ln(r), \;\;\;\;   \tau = t,
\eq
or inversely
\bq
\lb{B.2}
t = \tau,\;\;\;\;
r = e^{x + \tau}. 
\eq
Then, for any given function $f(t,\; r)$ we find that
\bqn
\lb{B.3}
f_{,t} &=&  f_{,\tau} - f_{,x},\;\;\;\; \;\;\;\;
f_{,r} =  \frac{1}{r} f_{,x},\nb\\
f_{,tr} &=& - \frac{1}{r}
\left(f_{,xx} - f_{,\tau x}\right),\;\;\;  
f_{,rr} = \frac{1}{r^{2}}
\left(f_{,xx} - f_{,x}\right),\nb\\   
f_{,tt} &=& f_{,\tau\tau} - 2 f_{,\tau x} + f_{,xx}.
\eqn
Substituting these expressions into
Eqs.(\ref{A.3a})-(\ref{A.3d}), we find that 
\bqn
\lb{B.4a}
G_{tt} &=& - \frac{e^{-2\Psi}}{r^{2}S^{2}}\left\{e^{2\Phi}\left[2SS_{,xx} +
          S_{,x}\left(S_{,x} + 4S\right) - 2S\Psi_{,x} \left(S_{,x} + S\right)
          + S^{2} - e^{2\Psi}\right]\right.\nb\\
& & \left. - r^{2}e^{2\Psi}\left(S_{,\tau} -
         S_{,x}\right)\left[2S\left(\Psi_{,\tau} - \Psi_{,x}\right) 
         + \left(S_{,\tau} - S_{,x}\right)\right]\right\},\\
\lb{B.4b}
G_{tr} &=& \frac{2}{rS}\left[S_{,xx} - S_{,\tau x} + 
           \left(S_{,x} + S\right)\left(\Psi_{,\tau} - \Psi_{,x}\right) 
         + \left(S_{,\tau} - S_{,x}\right)
           \left(\Phi_{,x} - 1\right)\right],\\
\lb{B.4c}
G_{rr} &=& \frac{e^{-2\Phi}}{r^{2}S^{2}}\left\{e^{2\Phi}\left[2S\Phi_{,x} 
            \left(S_{,x} + S\right) + S_{,x} \left(S_{,x} + 2S\right)   
       + S^{2} - e^{2\Psi}\right]\right.\nb\\ 
& &   - r^{2}e^{2\Psi}\left[2S\left(S_{,\tau\tau} - 2S_{,\tau x} +
      S_{,xx}\right)  \right.\nb\\
& & \left.\left. +   \left(S_{,\tau} - S_{,x}\right) \left(S_{,\tau} -
     2S\Phi_{,\tau} - S_{,x} + 2S\Phi_{,x}\right)\right]\right\},\\
\lb{B.4d}
G_{\theta\theta} &=& Se^{-2(\Phi + \Psi)} \left\{e^{2\Phi}\left[S\Phi_{,xx}  
          + S_{,xx} + \left(\Phi_{,x} - \Psi_{,x}\right)\left(S\Phi_{,x} +
         S_{,x} + S\right) + S_{,x} - S\Phi_{,x}\right]\right.\nb\\
  & &   - r^{2}e^{2\Psi}\left[S\Psi_{,xx} + S_{,xx} + S\left(\Psi_{,\tau\tau}
        - 2\Psi_{,\tau x}\right) + \left(S_{,\tau \tau} - 
          2S_{,\tau x}\right)\right.\nb\\
   & & \left.\left. - \left(S\Psi_{,\tau} + S_{,\tau} - S\Psi_{,x} -
        S_{,x}\right)\left(\Phi_{,\tau}   - \Psi_{,\tau} - \Phi_{,x}   
        + \Psi_{,x}\right)\right]\right\}.
\eqn
For the solutions with self-similarity of the zeroth kind, the metric
coefficients $\Phi,\; \Psi$ and $S$ are  functions of $x$ only,
\bq
\lb{B.5}
\Phi(\tau, x) = \Phi(x),\;\;\;\;
\Psi(\tau, x) = \Phi(x),\;\;\;\;
S(\tau, x) = S(x).
\eq
Then, all the derivatives of these functions with respect to
$\tau$ are zero, and Eqs.(\ref{B.4a}) - (\ref{B.4d}) reduce to,
\bqn
\lb{B.5a}
G_{tt} &=& \frac{e^{-2\Psi}}{r^{2}}\left\{e^{2\Phi}\left[2y_{,x} 
            - 2(1+y)\Psi_{,x} + 3y^{2} + 4y + 1 -  S^{-2}e^{2\Psi}\right]
           \right.\nb\\ 
& & \left. - r^{2}e^{2\Psi}y\left(2\Psi_{,x}+ y\right)\right\},\\
\lb{B.5b}
G_{tr} &=& \frac{2}{r}\left[y_{,x} - (1+y)\left(\Psi_{,x} - y\right) 
           - y\Phi_{,x}\right],\\
\lb{B.5c}
G_{rr} &=& \frac{e^{-2\Phi}}{r^{2}}\left\{e^{2\Phi}\left[
(1+y)\left(2\Phi_{,x} + y + 1\right) -   S^{-2}e^{2\Psi}\right]\right.\nb\\ 
& &  \left. - r^{2}e^{2\Psi}\left(2y_{,x} - 2y \Phi_{,x} + 3y^{2}\right)
 \right\},\\
\lb{B.5d}
G_{\theta\theta} &=& S^{2}e^{-2(\Phi + \Psi)}
\left\{e^{2\Phi}\left[\Phi_{,xx} +y_{,x} + \Phi_{,x}\left(\Phi_{,x} -
\Psi_{,x} + y\right) - (1+y)\left(\Psi_{,x} - y\right)\right]
\right.\nb\\
   & &  \left. - r^{2}e^{2\Psi}\left[\Psi_{,xx} + y_{,x} +
\left(\Psi_{,x}  -\Phi_{,x}\right)\left(\Psi_{,x} + y\right) 
+ y^{2}\right]\right\},
\eqn
where
\bq
\lb{B.6}
y = \frac{S_{,x}}{S}.
\eq

For a perfect fluid, the energy-momentum tensor (EMT) takes the form,
\bq
\lb{B.7}
T_{\mu\nu} = (\rho + p)u_{\mu}u_{\nu} - p g_{\mu\nu},
\eq
where, when the fluid is co-moving with the frame of the coordinates, its
four-velocity, $u_{\mu}$, is given by
\bq
\lb{B.8}
u_{\mu} = r_{1}e^{\Phi}\delta_{\mu}^{t}.
\eq
Then, from  the
$01$-component of the  Einstein field equations 
$
G_{\mu\nu} = T_{\mu\nu}
$,
and  Eq.(\ref{B.5b}) we find that  
\bq
\lb{B.9}
y_{,x} - (1+y)\left(\Psi_{,x} - y\right) - y \Phi_{,x} = 0, \;\;\;\; 
\left(G_{01} = T_{01}\right),
\eq
while the other components yield
\bqn
\lb{B.10a}
\rho &=& \frac{1}{\left(r_{1}r\right)^{2}}
\left\{e^{-2\Psi}\left(2y\Phi_{,x} + (1+y)^{2} -  
S^{-2}e^{2\Psi}\right]\right.\nb\\
& & \left. - r^{2}e^{-2\Phi}\left(2\Psi_{,x} + y\right)y\right\}, \;\;\;\; 
\left(G_{00} = T_{00}\right), \\   
\lb{B.10b}
p &=& \frac{1}{\left(r_{1}r\right)^{2}} 
\left\{e^{-2\Psi}\left[(1 + y)\left(2\Phi_{,x}+ y + 1\right)   
- S^{-2}e^{2\Psi}\right]\right.\nb\\ 
& & \left. - r^{2}e^{-2\Phi}\left[2(1+y)\Psi_{,x} + y(y - 2)\right]\right\},
\;\;\;\; \left(G_{11} = T_{11}\right),\\  
\lb{B.10c}
p &=& \frac{1}{\left(r_{1}r\right)^{2}} 
\left\{e^{-2\Psi}\left[\Phi_{,xx}+\Phi_{,x}\left(\Phi_{,x} - \Psi_{,x} +
2y\right)\right]\right.\nb\\
& &   \left. - r^{2}e^{-2\Phi} \left[\Psi_{,xx}+\Psi_{,x}\left(\Psi_{,x} -
\Phi_{,x} + 2y + 1\right)- y\right]\right\},
\;\;\;\; \left(G_{22} = T_{22}\right).
\eqn
Note that in writing the above equations, Eq.(\ref{B.9}) was used.  To have
Eqs.(\ref{B.10b}) and (\ref{B.10c}) be consistent, we must have
\bqn
\lb{B.11a}
\Psi_{,xx} + \Psi_{,x}\left(\Psi_{,x} - \Phi_{,x} - 1\right)  - y(y -1)  &=&
0,\\ 
\lb{B.11b} 
\Phi_{,xx} + \Phi_{,x}\left(\Phi_{,x} - \Psi_{,x} - 2\right)
- (1 + y)^{2}  + S^{-2}e^{2\Psi} &=& 0,\; (\alpha = 0).
\eqn

\subsection{Solutions with Self-Similarity of the Second Kind}

To study solutions with kinematic self-similarity of the second kind,  let us
 introduce other two new dimensionless variables, $x$ and $\tau$, via the
relations  
\bq
\lb{C.1a}
x = \ln\left[\frac{r}{(- t)^{1/\alpha}}\right],\;\;\;\;
\tau = - \ln\left(-t \right), 
\eq
or inversely
\bq
\lb{C.1b}  
t = - e^{-\tau},\;\;\;\;
r=  e^{x - \tau/\alpha},
\eq
where   $\alpha$ is a {\em dimensionless} constant. Then,  for any given
function  $f(t,r)$ we find that
\bqn
\lb{C.2}
f_{,t} &=& - \frac{1}{\alpha t} \left(\alpha f_{,\tau} +
f_{,x}\right),\;\;\;\; \;\;\;\;
f_{,r} =  \frac{1}{r} f_{,x},\nb\\
f_{,tr} &=& - \frac{1}{\alpha t r}
\left(f_{,xx} + \alpha f_{,\tau x}\right),\;\;\;  
f_{,rr} = \frac{1}{r^{2}}
\left(f_{,xx} - f_{,x}\right),\nb\\   
f_{,tt} &=&
\frac{1}{\alpha^{2}t^{2}}\left(\alpha^{2}f_{,\tau\tau} +
2\alpha f_{,\tau x} + f_{,xx} + \alpha^{2}f_{,\tau} + \alpha
f_{,x}\right).
\eqn
Substituting these expressions into
Eqs.(\ref{A.3a})-(\ref{A.3d}), we find that 
\bqn 
\lb{C.3a}
G_{tt} &=& - \frac{e^{-2\Psi}}{\alpha^{2}r^{2}S^{2}}\left\{
\alpha^{2}e^{2\Phi}\left[2SS_{,xx} + S_{,x}\left(S_{,x} + 4S\right)
-2S\Psi_{,x}\left(S_{,x} + S\right) + S^{2} - e^{2\Psi}\right]\right.\nb\\
& & \;\;\;\; - \frac{r^{2}}{t^{2}} e^{2\Psi} \left(2S\Psi_{,x} +
S_{,x}\right)S_{,x}\nb\\ 
& & \left. -  \frac{\alpha r^{2}}{t^{2}} e^{2\Psi} \left[ 2S\left(\alpha
S_{,\tau}\Psi_{,\tau} + S_{,\tau}\Psi_{,x} + \Psi_{,\tau}S_{,x}\right) +
S_{,\tau}\left(\alpha S_{,\tau}  + 2S_{,x}\right)\right]\right\},\\
\lb{C.3b}
G_{tr} &=&  \frac{2 }{\alpha r t S}\left\{
S_{,xx} - \Psi_{,x}\left(S_{,x} + S\right) - S_{,x}\left(\Phi_{,x} -
1\right)\right.\nb\\ 
& & \left. + \alpha \left[S_{,\tau x} - \Psi_{,\tau}\left(S_{,x} + S\right)
- S_{,\tau}\left(\Phi_{,x} - 1\right)\right]\right\},\\
\lb{C.3c}
G_{rr} &=& \frac{e^{-2\Phi}}{\alpha^{2}r^{2}S^{2}}\left\{ 
\alpha^{2}e^{2\Phi}\left[2S\Phi_{,x}\left(S_{,x} + S\right)
+ S_{,x}\left(S_{,x} + 2S\right) + S^{2} - e^{2\Psi}\right]\right.\nb\\
& & \;\;\;\; - \frac{r^{2}}{t^{2}} e^{2\Psi} \left[2SS_{,xx} +
S_{,x}\left(S_{,x} - 2S\Phi_{,x} + 2\alpha S\right) \right]\nb\\ 
& &  - \frac{\alpha r^{2}}{t^{2}} e^{2\Psi} \left[ 2S\left(\alpha
S_{,\tau\tau} + 2 S_{,\tau x}\right) + S_{,x}\left(S_{,\tau} -
2S\Phi_{,\tau}\right)\right.\nb\\ 
& & \left.\left. \;\;\; +
S_{,\tau}\left(\alpha S_{,\tau} - 2\alpha S\Phi_{,\tau} + S_{,x} - 2S\Phi_{,x}
+ 2\alpha S\right)\right]\right\},\\ 
\lb{C.3d}
G_{\theta\theta} &=& \frac{S}{\alpha^{2}}\left\{ 
\alpha^{2}e^{-2\Psi}\left[S\Phi_{,xx} + S_{,xx} + \left(\Phi_{,x}
- \Psi_{,x}\right)\left(S\Phi_{,x} + S_{,x} + S\right) - S\Phi_{,x} +
S_{,x}\right]\right.\nb\\ 
& &  - \frac{r^{2}}{t^{2}} e^{-2\Phi}\left[S\Psi_{,xx} + S_{,xx} -
\left(S\Psi_{,x} + S_{,x}\right)\left(\Phi_{,x} - \Psi_{,x} -
\alpha\right)\right]\nb\\ 
& & - \frac{\alpha r^{2}}{t^{2}} e^{-2\Phi} \left[ S\left(\alpha
\Psi_{,\tau\tau} + 2 \Psi_{,\tau x}\right) + \alpha S_{,\tau\tau} + 2 S_{,\tau
x} \right.\nb\\ 
& & - \left(S\Psi_{,\tau}+S_{,\tau}\right)\left(\alpha\Phi_{,\tau} -
\alpha\Psi_{,\tau} + \Phi_{,x} - \Psi_{,x} - \alpha\right)\nb\\ & &
\left.\left. - \left(\Phi_{,\tau}- \Psi_{,\tau}\right)\left(S\Psi_{,x} +
S_{,x}\right)\right]\right\}.
\eqn
 
For the solutions with self-similarity of the second kind, the metric
coefficients are also functions of $x$ only, but now with $x$ being given
by Eq.(\ref{C.1a}). Then, setting all the derivatives with respect to $\tau$
zero, Eqs.(\ref{C.3a}) - (\ref{C.3d}) reduce to 
\bqn 
\lb{C.4a}
G_{tt} &=& - \frac{1}{r^{2}}e^{2(\Phi -\Psi)} \left[2y_{,x} + y(3y +
4) + 1 - 2(1+y)\Psi_{,x} - S^{-2}e^{2\Psi}\right]\nb\\
& & + \frac{1}{\alpha^{2}t^{2}}(2\Psi_{,x} + y)y,\\
\lb{C.4b}
G_{tr} &=&  \frac{2}{\alpha t r} 
\left[y_{,x} + (1+y)(y - \Psi_{,x}) - y\Phi_{,x}\right],\\
\lb{C.4c}
G_{rr} &=& \frac{1}{r^{2}}\left[
2(1+y)\Phi_{,x} + (1 + y)^{2}  - S^{-2}e^{2\Psi}\right]\nb\\
& & - \frac{1}{\alpha^{2}t^{2}}e^{2(\Psi - \Phi)}
\left[2y_{,x} + y\left(3y - 2\Phi_{,x} + 2\alpha\right)\right],\\
\lb{C.4d}
G_{\theta\theta} &=&  S^{2}e^{-2\Psi}\left[\Phi_{,xx} + y_{,x} +
\Phi_{,x}\left(\Phi_{,x} - \Psi_{,x} + y\right) +
\left(1 + y\right)\left(y - \Psi_{,x}\right)\right]\nb\\ 
& &  - \frac{r^{2}S^{2}}{\alpha^{2}t^{2}}e^{-2\Phi}\left[\Psi_{,xx} + y_{,x} +
y^{2} - \left(\Psi_{,x} + y\right)\left(\Phi_{,x} - \Psi_{,x} -
\alpha\right)\right]. 
\eqn

For a perfect fluid, the  EMT is given by Eqs.(\ref{B.7}) and (\ref{B.8}).
Similarly, from  the
$01$-component of the  Einstein field equations 
we find that  
\bq
\lb{C.5}
y_{,x} - (1+y)\left(\Psi_{,x} - y\right) - y \Phi_{,x} = 0, \;\;\;\; 
\left(G_{01} = T_{01}\right),
\eq
while the other components yield
\bqn
\lb{C.6a}
\rho &=& \frac{1}{\kappa r^{2}_{1}}
\left\{\frac{1}{\alpha^{2}t^{2}}e^{-2\Phi}\left(2\Psi_{,x} + y\right)y
\right.\nb\\ 
& & \left. - \frac{1}{r^{2}}e^{-2\Psi}\left[2y\Phi_{,x} + (1 + y)^{2} -
S^{-2}e^{2\Psi}\right]\right\}, \;\;\;\;  \left(G_{00} =  T_{00}\right),
\\   
\lb{C.6b}
p &=& \frac{1}{\kappa r^{2}_{1}}
\left\{\frac{1}{r^{2}}e^{-2\Psi}\left[2(1 + y)\Phi_{,x} + (1 +
y)^{2} - S^{-2}e^{2\Psi}\right]\right.\nb\\
& & \left. - \frac{1}{\alpha^{2}t^{2}} e^{-2\Phi}\left[2(1 + y)\Psi_{,x} +
y^{2} + 2(\alpha - 1)y\right]\right\},\;\;\;\; \left(G_{11} = 
T_{11}\right),\\  
\lb{C.6c}
p &=& \frac{1}{\kappa r^{2}_{1}}
\left\{\frac{1}{r^{2}}e^{-2\Psi}\left[\Phi_{,xx} +  
\Phi_{,x}\left(\Phi_{,x} - \Psi_{,x} + 2y\right)\right]\right.\nb\\
& & \left. - \frac{1}{\alpha^{2}t^{2}} e^{-2\Phi}\left[\Psi_{,xx} + 
\Psi_{,x}\left(\Psi_{,x} - \Phi_{,x} + 2y + \alpha + 1\right)
+ (\alpha - 1)y\right]\right\}, \;\;\;\; \left(G_{22} =  T_{22}\right),
\eqn 
where in writing the above equations, Eq.(\ref{C.5}) was used.

When $\alpha \not= 1$, in the expressions of $p$ the term   that is
proportional to $r^{-2}$ has different power-dependence on $r$ from the term
that is  proportional to $t^{-2}$, when these expressions are written in
terms of $r$ and $x$, since 
\bq
\lb{C.7}
t = - r^{\alpha}e^{-\alpha x}.
\eq
Then, the two expressions of Eqs.(\ref{C.6b}) and (\ref{C.6c}) are equal
only when
\bqn
\lb{C.8a}
\Psi_{,xx} + \Psi_{,x}\left(\Psi_{,x} - \Phi_{,x}\right) + (\alpha
-1)\left(\Psi_{,x} - y\right) - y^{2}   &=& 0,\\
\lb{C.8b}
\Phi_{,xx} + \Phi_{,x}\left(\Phi_{,x} - \Psi_{,x} - 2\right) - (1 +
y)^{2}  + S^{-2}e^{2\Psi} &=& 0.
\eqn

When $\alpha = 1$,  all the terms  in the expressions of $p$ have the same
power-dependence on $r$,  and the two expressions of Eqs.(\ref{C.6b}) and
(\ref{C.6c}) are equal, provided that 
\bqn
\lb{C.9}
& & \Phi_{,xx} + \Phi_{,x}\left(\Phi_{,x} - \Psi_{,x} - 2\right) - (1 +
y)^{2}  + S^{-2}e^{2\Psi} \nb\\
& & - e^{2(x + \Psi - \Phi)}
\left[\Psi_{,xx} + \Psi_{,x}\left(\Psi_{,x} - \Phi_{,x}\right) - y^{2}\right]
= 0,\;\; (\alpha = 1).
\eqn
From the above equations we can see that a
solution  that satisfies  Eqs.(\ref{C.8a}) and (\ref{C.8b}) with $\alpha =
1$ is also a solution of Eq.(\ref{C.9}), but not the other way around, that
is, a solution of Eq.(\ref{C.9}) doesn't necessarily satisfy
Eqs.(\ref{C.8a}) and (\ref{C.8b}).


\begin{thebibliography}{100}


\bibitem{Joshi93} P.S. Joshi, {\em Global Aspects in Gravitation 
and Cosmology} (Clarendon, Oxford, 1993).

\bibitem{Ch93} M.W. Choptuik, Phys. Rev. Lett. {\bf 70}, 9 (1993).

\bibitem{Gu00} C. Gundlach, Adv. Theor. Math. Phys. {\bf
2}, 1 (1998), {\tt gr-qc/9712084} (1997); ``{\em Critical Phenomena in
Gravitational Collapse}," {\tt gr-qc/0001046}, (2000).

\bibitem{Wang01} A.Z. Wang, ``{\em Critical Phenomena in Gravitational
Collapse: The Studies So Far}", {\tt gr-qc/0104073}, to appear in Braz. J.
Phys. (2001).

\bibitem{HKA96} T. Hara, T. Koike, and S. Adachi, ``{\em Renormalization 
group and critical behavior in gravitational collapse},"  
$gr-qc/9607010$, preprint (1996); T. Koike, T. Hara, and S. Adachi, Phys. Rev.
{\bf D59},  104008 (1999).

\bibitem{BCG97} P.R. Brady, C.M. Chambers, and S.M.C.V. Gon\c{c}alves, Phys.
Rev. {\bf D56}, 6057 (1997).

\bibitem{EC94} C.R. Evans and J.S. Coleman, Phys. Rev. Lett. {\bf 72}, 
1782 (1994); T. Koike, T. Hara, and S. Adachi, {\em ibid.} {\bf74}, 5170
(1995); D. Maison, Phys. Lett. {\bf B366}, 82 (1996).

\bibitem{CH89} B. Carter and R.N. Henriksen, Ann. Physique Suppl. {\bf 14}, 47
(1989).

\bibitem{BC98} P.M. Benoit and A.A. Coley, Class. Quantum Grav. {\bf 15},
2397  (1998); P.M. Benoit, ``{\em Qualitative Investigations of
Solutions to Einstein's Field equations Admitting a Kinematic
Self-Similarity}," Ph.D. thesis, Dalhousie University, Halifax, Nova Scotia,
Canada (1999).


\bibitem{BZ72} G.E. Barenblatt and Ya.B. Zel'dovich, Ann. Rev. Fluid. Mech.
{\bf 4}, 285 (1972).

\bibitem{CT71} A.H. Cahill and M.E. Taub, Commu. Math. Phys. {\bf 21}, 1
(1971).

\bibitem {Coley97} A.A. Coley, Class. Quantum Grav. {\bf 14}, 87 (1997).


\bibitem{TB34} R.C. Tolmann, Proc. Natl. Acad. Sci. (USA), {\bf 20}, 169
(1934); H. Bondi, Mon. Not. Astron. Soc. {\bf 107}, 400 (1948).

\bibitem{PI90}E. Poisson and W. Israel,  Phys. Rev. {\bf D41}, 1796 (1990).

\bibitem{WO97} A.Z. Wang and H.P. de Oliveira, Phys. Rev. {\bf D56}, 753
(1997).

\bibitem{WRS97} A.Z. Wang, J.F.Villas da Rocha, and N.O. Santos, 
 Phys. Rev. {\bf D56}, 7692 (1997); J.F.Villas da Rocha, A.Z. Wang, and
N.O. Santos, Phys. Lett. {\bf A255}, 213 (1999); J.F.Villas da Rocha and A.Z.
Wang, Class. Quantum Grav. {\bf 17}, 2589 (2000).

\bibitem{HE73} S.W. Hawking and G.F.R. Ellis, {\em The Large Scale
Structure of Spacetime}, (Cambridge University Press, Cambridge,
1973).

\bibitem{LL75} L.D. Landau and E.M. Lifshitz, {\em The Classical Theory of
Fields}, (Pergamon Press, New York, 1975), pp. 309-312.



 \end{thebibliography}
\end{document}